\documentclass[runningheads,citeauthoryear]{apinv}
\usepackage{epsfig,cite,graphics}
\usepackage{marvosym} 
\usepackage[utf8]{inputenc}
\usepackage{graphicx}
\usepackage{xcolor}
\usepackage{subfigure}
\usepackage[hidelinks]{hyperref}
\usepackage{multirow}
\usepackage{amsmath}
\usepackage{breqn}
\usepackage{amssymb}
\usepackage{url}

\begin{document}

\title{Fundamentals of Stellar Parameters Estimation \\
through CMD of Star Clusters: Open (NGC \\
2360) and Globular (NGC 5272)
}
\titlerunning{Fundamentals of Stellar Parameters Estimation}
\author{Kanwar Preet Kaur\inst{}, Pankaj S. Joshi\inst{}}
\authorrunning{KP Kaur, PS Joshi}
\tocauthor{Kanwar Preet Kaur, Pankaj S. Joshi} 
\institute{International Center for Cosmology, Charotar University of Science and Technology, Gujarat 388421, India
   \newline
	\email{kanwarpreet27@gmail.com}    }
\papertype{Submitted on xx.xx.xxxx; Accepted on xx.xx.xxxx}	
\maketitle

\begin{abstract}

The fundamentals of estimating essential stellar parameters of an open cluster - NGC 2360 and globular clusters - NGC 5272 are presented extensively in this work. Here, the evaluation of stellar parameters, by manually fitting the appropriate isochrones on the color magnitude diagrams (CMDs), of the selected star clusters is discussed comprehensively. Aperture photometry and point-spread function fitting (PSF) photometry are conducted on \textit{g}, \textit{r}, and \textit{i} standard band filter images of Sloan Digital Sky Survey (SDSS) using the aperture photometry tool (APT) to obtain the respective CMDs. Further, to achieve the stellar parameters, isochrone fitting is described in detail. This work on stellar parameters evaluation has attained the following results: age of NGC 2360 is found to be 708 Myrs with metallicity, [Fe/H], of -0.15, whereas NGC 5272 is having age of 11.56 Gyrs with metallicity, [Fe/H], of -1.57. Additionally, the interstellar reddening, $E(B-V)$, and distance modulus, $DM$, for NGC 2360 are obtained as 0.12 and 11.65, respectively. While, for NGC 5272, the interstellar reddening is attained as $E(B-V)$=0.015, and the distance modulus is $DM$=15.1. The values of these stellar parameters are found to be in close approximation with the results provided in the literature based on the IRAF analysis technique. The distribution of radii, masses, and temperatures are included along with the initial mass function (IMF) for both the start clusters. Thus, this article would aid in providing insight into the evaluation of stellar parameters by the astronomical photometry analysis which would successively upsurge the understanding of our universe. However, it should be noted that the cleaning of cluster population on the CMDs from the foreground\slash background stars, clearing of spurious objects, error estimations and the membership determination are not carried out in this work and are considered as separate project for analysis.
\end{abstract}
\keywords{Stellar Parameters, Open Cluster, Globular Cluster, Astronomical Photometry}

\section*{Introduction}
\label{sec:Intro}

{For understanding the stellar evolution and formation of stars, it is essential to study the star clusters. Generally, a color magnitude diagram (CMD) is plotted and analysed for a given star cluster \cite{Allison} to provide various significant stellar parameters by fitting the optimum theoretical isochrones. CMD itself provides the summary of the magnitude difference (color) and magnitude of a celestial object. Magnitude is a measure of brightness and is obtained by performing astronomical photometry on different color band images of a celestial object. In this work, an open cluster NGC 2360 and a globular cluster NGC 5272 are studied extensively to provide the fundamentals of stellar parameter estimations. McClure \cite{McClure} has reported the photoelectric photometry of the giant stars in NGC 2360 along with its photometric properties. Then G\"{u}ne\c{s} with his team \cite{Gunecs} has analysed the NGC 2360 to derive the astrophysical and structural parameters using 2MASS filter data. Here, the author has depicted that 2MASS JHKs $E(B-V)$ reddening values agrees with the values derived from UBV and $ubvy-\beta$ photometry. Single red giants of the NGC 2360 are analysed by Su\'{a}rez and group \cite{Suarez} to provide the atmospheric parameters, abundances of the elements, radial, and rotational velocities. The stellar parameter estimation from UBVR$_{c}$I$_{c}$ photometric analysis of NGC 2360 is presented by Maurya and Joshi \cite{Maurya}. They provided data for mean reddening of $E(B-V)$, age, distance, and mass function of NGC 2360 along with the examination of the cluster's membership. The recent updates on NGC 2360 could be obtained from the article of Cantat-Gaudin and Anders \cite{Cantat-Gaudin}. For the thorough color-color transformation (Gaia to UBVR$_{c}$I$_{c}$) one could refer \cite{Riello, Ritter}.

The globular cluster NGC 5272 is also a well-researched topic. Preliminary results of CCD photometry analysis of NGC 5272 are presented by Paez and his team \cite{Paez}. Further, Buonanno with his team \cite{Buonanno} have provided an in-depth analysis and study of NGC 5272. Here, the author has studied various sections of NGC 5272 CMD in detail. D. An and the group \cite{An} have provided the analysis of cluster photometry and fiducial sequences for NGC 5272 using Sloan Digital Sky Survey (SDSS) imaging data, whereas Str{\"o}mgren photometry is presented by Massari and group \cite{Massari}. The age and temperature of NGC 5272 are obtained from the CCD Photometry analysis by Thassana and Maithong \cite{Thassana}. Dodd with his team \cite{Dodd} have utilized the data taken from WFC/INT and the \textit{Hubble Space Telescope (HST)} for the photometric analysis of NGC 5272 to obtain Hertzsprung-Russell(HR) diagrams. The HR diagram is fitted with the optimum isochrone to evaluate the age, distance, metallicity, and reddening of the cluster. Moreover, the latest updates on NGC 5272 could be found in \cite{Baumgardt}.

Most of the reported astronomical photometry articles in the existing literature which are based on the estimation of the stellar parameter are usually performed using either ``IRAF Data Reduction and Analysis System" \cite{Massey}, or a Linux based analysis procedure, or using Python-based package, for instance, ``ASTROPY" \cite{Robitaille}. For amateur astronomers, these prevailing techniques could be difficult to understand and perform. Hence, this work presents the photometry analysis of NGC 2360 (open cluster) and NGC 5272 (globular cluster) by utilizing a Windows-based software, viz. aperture photometry tool (APT) \cite{Laher}. These well-researched open clusters and globular clusters are selected so that the manually obtained results could be compared with the literature results along with the understanding of the physics behind isochrone fitting and stellar parameter estimation. 

The photometry analysis results obtained from APT are then utilized to evaluate some of the significant stellar parameters values. Although, exact values of the stellar parameters are difficult to obtain from only photometric data. However, the following reasonable values for the stellar parameters are evaluated using photometry analysis by APT software: age of NGC 2360 is found to be 708 Myrs with metallicity, [Fe/H], of -0.15, whereas NGC 5272 is having age of 11.56 Gyrs with metallicity, [Fe/H], of -1.57. In addition, the interstellar reddening, $E(B-V)$, and distance modulus, $DM$, for NGC 2360 are obtained as 0.12 and 11.65, respectively. While, for NGC 5272, the interstellar reddening is attained as $E(B-V)$=0.015, and the distance modulus is $DM$=15.1. These stellar parameter values are found to be in close approximation with the results reported in the literature which are based on the IRAF analysis technique (Table \ref{t2}). In addition to these parameters absorption due to interstellar extinction, temperature, surface brightness, angular radius, and luminosity of the sources present in the two chosen star clusters are evaluated along with their respective initial mass functions (IMF). Thus, this work would notably help amateur astronomers to familiarize themselves with the groundwork needed to perform astronomical photometry for estimating the stellar parameters from isochrones fitting on the CMDs. 

It should be noted that in this work all the necessary terminologies\slash techniques (to the best knowledge of the author) are mentioned in detail to perform the astronomical photometry along with the selection of photometry analysis (aperture or PSF), isochrones, and then to determine essential stellar parameters by fitting the appropriate isochrones. Thus, the cleaning of cluster population on the CMDs from the foreground\slash background stars and clearing of spurious detections are not performed herein. In addition, the error estimations and the membership determination are not carried out in this work. However, a detailed steps of fitting isochrone on the CMD and uncertainty estimation is presented in \hyperref[sec:SPE]{Section 4}. As these stated procedures, a significant measures, involves much more sophisticated processing that is beyond the extent of this work and would be considered as a separate project for analysis.

This paper is organized as follows: \hyperref[sec:SPE]{Section 1} provides brief details on astronomical photometry and the observational data information are provided in \hyperref[sec:SPE]{Section 2}. In \hyperref[sec:SPE]{Section 3}, the work description is presented. \hyperref[sec:SPE]{Section 4} describes the details on the stellar parameter estimation, while the presented work is summarized in Conclusion.}

\section*{1. Astronomical Photometry}
\label{sec:AP}

{Astronomical photometry \cite{Romanishin} which involves the study of astronomical objects by measuring their energy flux plays a significant role in determining various stellar parameters. In this work, two different star clusters are selected to perform the photometry techniques with an aim to provide details on the fundamentals of essential stellar parameters determination from their color magnitude diagrams (CMDs) and the theoretical isochrones. By evaluating stellar parameters one could observe the difference between open star clusters and globular star clusters which in turn would aid in understanding our universe to a great extent. These star clusters are selected because they are crucial for investigating the stellar evolution. For instance, investigation of the chosen star clusters would provide the knowledge of their age, metallicity, temperature, and distance, to name a few, which could help in exploring the evolution of our universe more proficiently.

The CMD is an alternate plot of Hertzsprung-Russell (HR) diagram or HRD. The HR diagram is plotted between celestial objects' temperature and their luminosity (magnitude), while CMD provides a summary of the color and magnitude of the celestial objects. In order to determine the magnitude of a star from a CCD image, it is essential to extract the total brightness of the respective star while excluding any brightness from the background sky. The star brightness is obtained by summing up the values stored in each pixel which are illuminated by both the star-light and the background sky-light. Conversely, the light contribution from the background sky is determined by adding the values of the nearby pixels. Then the total source-light is attained by subtracting net pixel values (star-light plus the background sky-light) from the nearby background sky-light pixel values. Generally, there are two approaches to determine the magnitude of a star or any celestial object, namely aperture photometry \cite{Berry} and point spread function photometry \cite{DaCosta, Heasley}. In aperture photometry, the fluxes from the celestial object and the background sky are gathered within a certain aperture. Usually, circular apertures are selected to extract the star-light and the background light. However, the aperture could be of any arbitrary shape. The star-light is extracted by adding up the pixel counts within circles centred on each object and subtracting off the sky count which is obtained from the annulus created around the star aperture at a certain distance. Aperture photometry is performed for the celestial objects which are neither too faint nor too crowded, for instance, open clusters. For faint and crowded star fields, for example globular clusters, this technique is not suitable because of the high stellar density that makes inaccurate average sky count determination, or sometimes it is even impossible to determine the average sky count. In order to overcome this limitation, the PSF photometry is applied to the globular clusters. PSF is a mathematical model of the stellar profile which describes the average distribution of the photons, on the CCD surface, coming from a single star. The light coming from a celestial object gets broadened due to diffraction caused by the earth's atmosphere. This light distribution is considered to be Gaussian, Moffat, Lorentz, or a combination of these distributions.

In view of this context, the all-sky aperture photometry is performed on the open cluster NGC 2360, whereas PSF fitting photometry is applied on the globular cluster NGC 5272 using APT software. The photometry analysis results are then utilized to evaluate some of the significant stellar parameter values. In spite of the fact that the precise values of the stellar parameters are tough to obtain from solely photometric data. Yet, reasonable values for the stellar parameters could be attained from the presented photometric data analysis. Thus, this work would provide a relatively good insight into estimating the age, metallicity, and distance in particular, of the selected star clusters from their respective CMDs. Besides the aforementioned parameters, the other parameters which would be evaluated using photometric data are distance modulus, interstellar reddening, absorption due to interstellar extinction, temperature, surface brightness, angular radius, luminosity, and IMF of the sources present in the two chosen star clusters. }

\section*{2. Observational Data}
\label{sec:OD}

{The color magnitude diagrams (CMDs) utilized in this work are obtained by performing aperture photometry and PSF photometry on NGC 2360 and NGC 5272, respectively. NGC 2360 is an open cluster located in the Canis Major constellation, whereas NGC 5272  is a  globular cluster situated in the Canes Venatici constellation. NGC 5272 is also popularly known as Messier 3 or M3. The celestial coordinates of NGC 2360 are: RA (J2000) = 07h 17m 43.1s  or 109.4297\textdegree~ and Dec (J2000) = -15\textdegree 38\arcmin 29\arcsec~ or -15.6413\textdegree. Further, for NGC 5272 the celestial coordinates are: RA (J2000) = 13h 42m 11.22s  or 205.5468\textdegree~ and Dec (J2000) = +28\textdegree 22\arcmin 31.6\arcsec~ or +28.3755\textdegree. These celestial coordinates are taken from NASA/IPAC Extragalactic Database (NED) \url{(https://ned.ipac.caltech.edu/)}. The images of both the selected star clusters are extracted from the Data Release 12 (DR12) of Sloan Digital Sky Survey (SDSS-III) \url{(https://dr12.sdss.org/)}. The \textit{r}-band images of the open cluster and the globular cluster are depicted in Fig. \,\ref{f1} having the size of 15\arcmin~ by 15\arcmin. The SDSS archive has images taken from a 2.5-m f/5 modified Ritchey-Chr\'{e}tien altitude-azimuth telescope located at Apache Point Observatory, in southeast New Mexico.

\begin{figure}[!htb]
  \begin{center}
  \subfigure[NGC 2360]
    {\centering{\includegraphics[width=60mm]{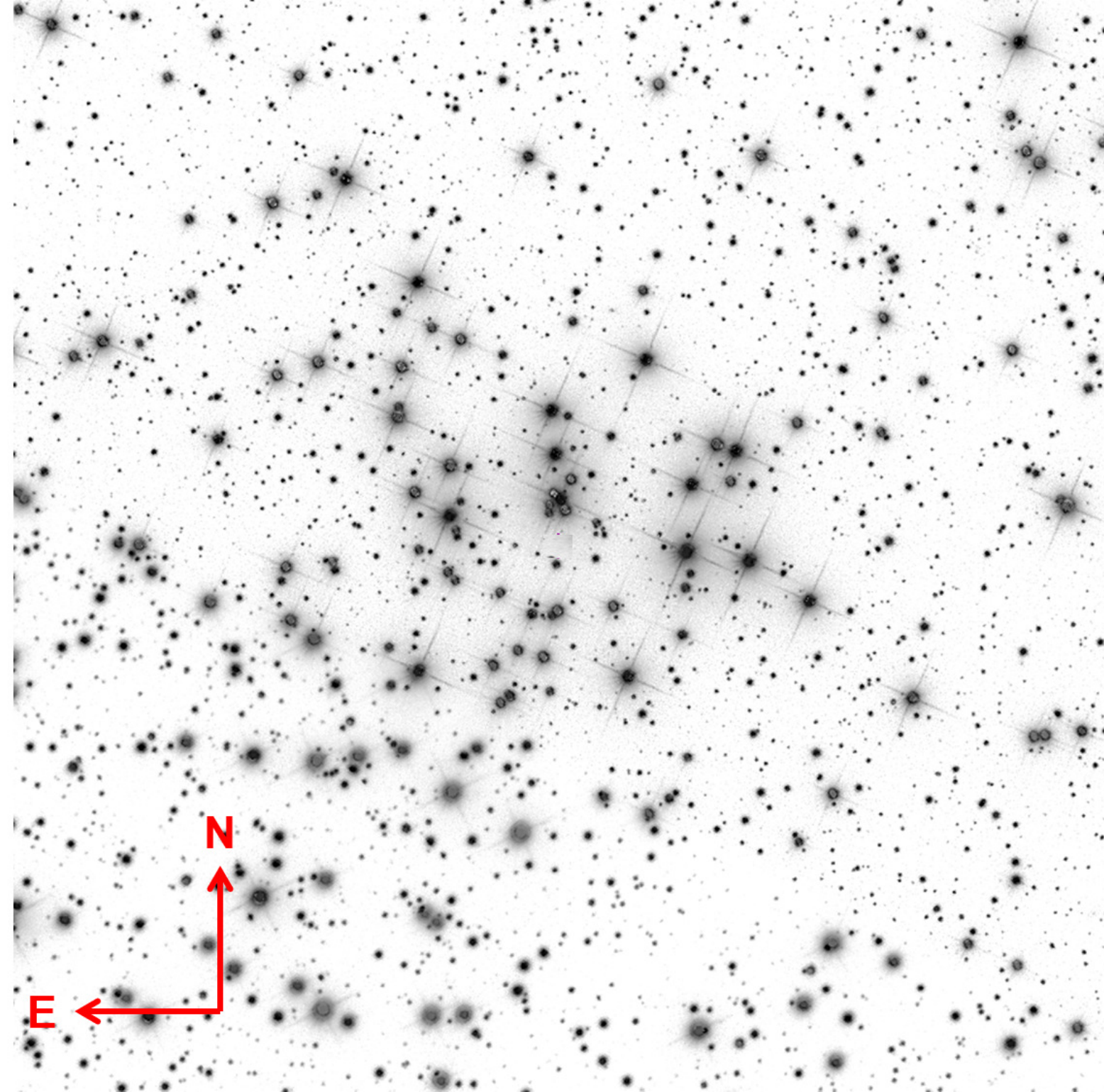}} \label{f1a}}
  \subfigure[NGC 5272]
    {\centering{\includegraphics[width=60mm]{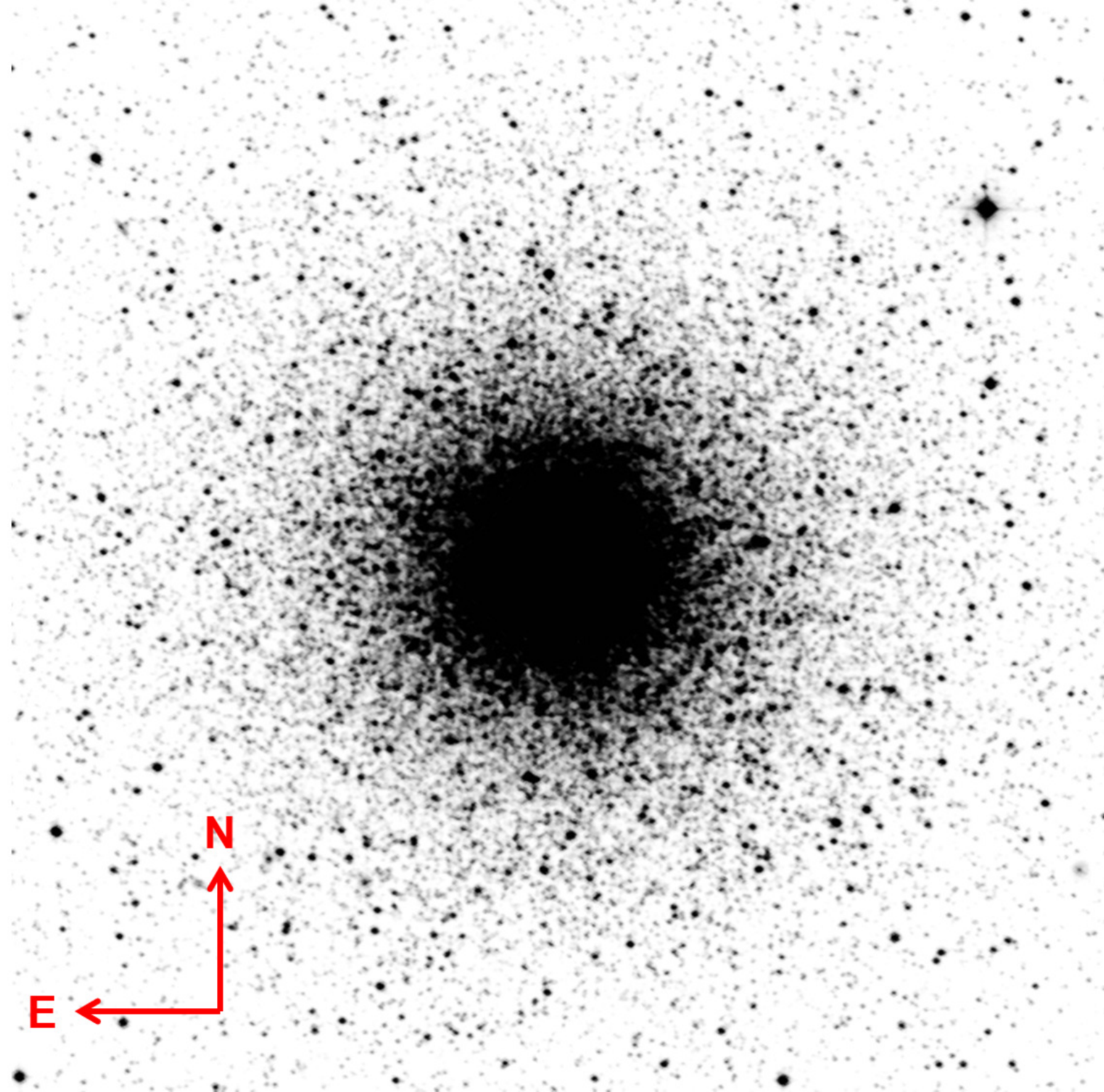}} \label{f1b}}
     \caption{\textit{r}-band images of: (a) open cluster and (b) globular cluster.} \label{f1}
  \end{center}
\end{figure}

For the photometry analysis, the calibrated images of star clusters in the three standard photometric bands \textit{g}, \textit{r}, and \textit{i} are utilized to obtain the instrumental magnitudes. The instrumental magnitude is then converted to a magnitude that is tied to the standard photometric system (standard or calibrated magnitude) by using the calibration parameters, viz. atmospheric extinction-coefficient, air-mass, zero point, color term, and color index. These photometric calibration parameters are extracted from the respective `tsfield' files of the star cluster images. }

\section*{3. Work Description}
\label{sec:WD}

{The all-sky photometry is performed on the selected star clusters using the aperture photometry tool (APT). The all-sky aperture as well as PSF photometry analysis are conducted on the \textit{g}, \textit{r}, and \textit{i} band images of NGC 2360 and NGC 5272 star clusters, respectively. The exposure time of these images is 53.907s. After the application of aperture photometry analysis on NGC 2360, APT has identified 1247, 1243, and 1244 photometrized objects in \textit{g}, \textit{r}, and \textit{i} color bands, respectively. While PSF photometry analysis of NGC 5272 using APT has yielded 3791 sources in \textit{g}, 3790 in \textit{r}, and 3791 in \textit{i} color bands. The source coordinates of both the star clusters are matched to obtain sources common to all the three bands. This source matching is performed in TOPCAT software which has ultimately identified 1243 and 3790 photometric objects in NGC 2360 and NGC 5272, respectively. The calibrated (standard) magnitudes ($M$) are obtained by using equation  Eq. (\ref{r1}) \cite{Palmer, Berry, Warner}. 

\begin{eqnarray}
 \label{r1}
M &=& m_{ins}-\Lambda+ZP+kX+C\times CI
\end{eqnarray}

In Eq. (\ref{r1}), $m_{inst}$ is the instrumental magnitude taking into account the effects of atmospheric extinction coefficient ($k$), air-mass($X$), color correction term ($C$), and color index ($CI$). $\Lambda$ and $ZP$ in the Eq. (\ref{r1}) above, represents the arbitrary constant and the zero point, respectively. The values of photometric calibration parameters are depicted in Table \ref{t1} which are extracted from the `tsfield' files presented in the SDSS archive. These photometric calibration parameters are utilized for the evaluation of calibrated magnitudes and corresponding magnitude errors for NGC 2360 and NGC 5272. More details on astronomical photometry to determine the calibrated magnitudes and the corresponding error magnitudes of stars could be found in \cite{Kaur}. After the calibration of \textit{g}, \textit{r}, and \textit{i} magnitudes, they are transformed into the \textit{B}, \textit{V}, and \textit{R} band magnitudes of Johnson-Cousins UBVRI standard photometric system by using transform equations: Eq. (\ref{r2}) to Eq. (\ref{r4}) \cite{Lupton}. The UBVRI standard photometric calibrated magnitudes are then employed to evaluate the stellar parameters of the chosen star clusters by plotting the graph between their apparent visual magnitudes ($V$) and the color indices ($B-V$): the color magnitude diagram (CMD).

\begin{eqnarray}
 \label{r2}
    B &=& g_{cal} + 0.3130\times{(g_{cal} - r_{cal})} + 0.2271
 \end{eqnarray}

\begin{eqnarray}
 \label{r3}
    V &=& g_{cal} - 0.5784\times{(g_{cal} - r_{cal})} - 0.0038
\end{eqnarray}

\begin{eqnarray}
 \label{r4}
    R &=& r_{cal} - 0.2936\times{(r_{cal} - i_{cal})} - 0.1439
\end{eqnarray}

\begin{table}[htbp]
  \centering
  \caption{Photometric calibration parameters}
    \begin{tabular}{ccc}
    \hline
    \multirow {2}{7em}{\centering{\textbf{Calibration Parameters}}} &      \multicolumn {2}{p{14em}} {\centering{\textbf{Selected Star Clusters}}} 
    \\
        \cline{2-3}   & 
    \multicolumn{1}{p{7em}}{\centering{\textbf{NGC 2360}}} & \multicolumn{1}{p{7em}}{\centering{\textbf{NGC 5272}}} \\
       \hline
    \textbf{\textit{X}} & 1.626 & 1.189  \\
    \textbf{\textit{k}} & 0.454 & 0.549  \\
    \textbf{\textit{ZP}} & -23.528 & -23.768  \\
    \hline
    \end{tabular}%
  \label{t1}%
\end{table}%

The CMD and error magnitude plots of NGC 2360 and NGC 5272 for the ugriz photometric system are depicted in Fig.\,\ref{f2} and Fig.\,\ref{f3}, respectively. Successively, the CMDs of NGC 2360 and NGC 5272 for the UBVRI standardized photometric system are illustrated in Figs.\,\ref{f4a} and \,\ref{f4b}, respectively.

\begin{figure}[!htb]
  \begin{center}
  \subfigure[Color magnitude diagram]
    {\centering{\includegraphics[width=60mm]{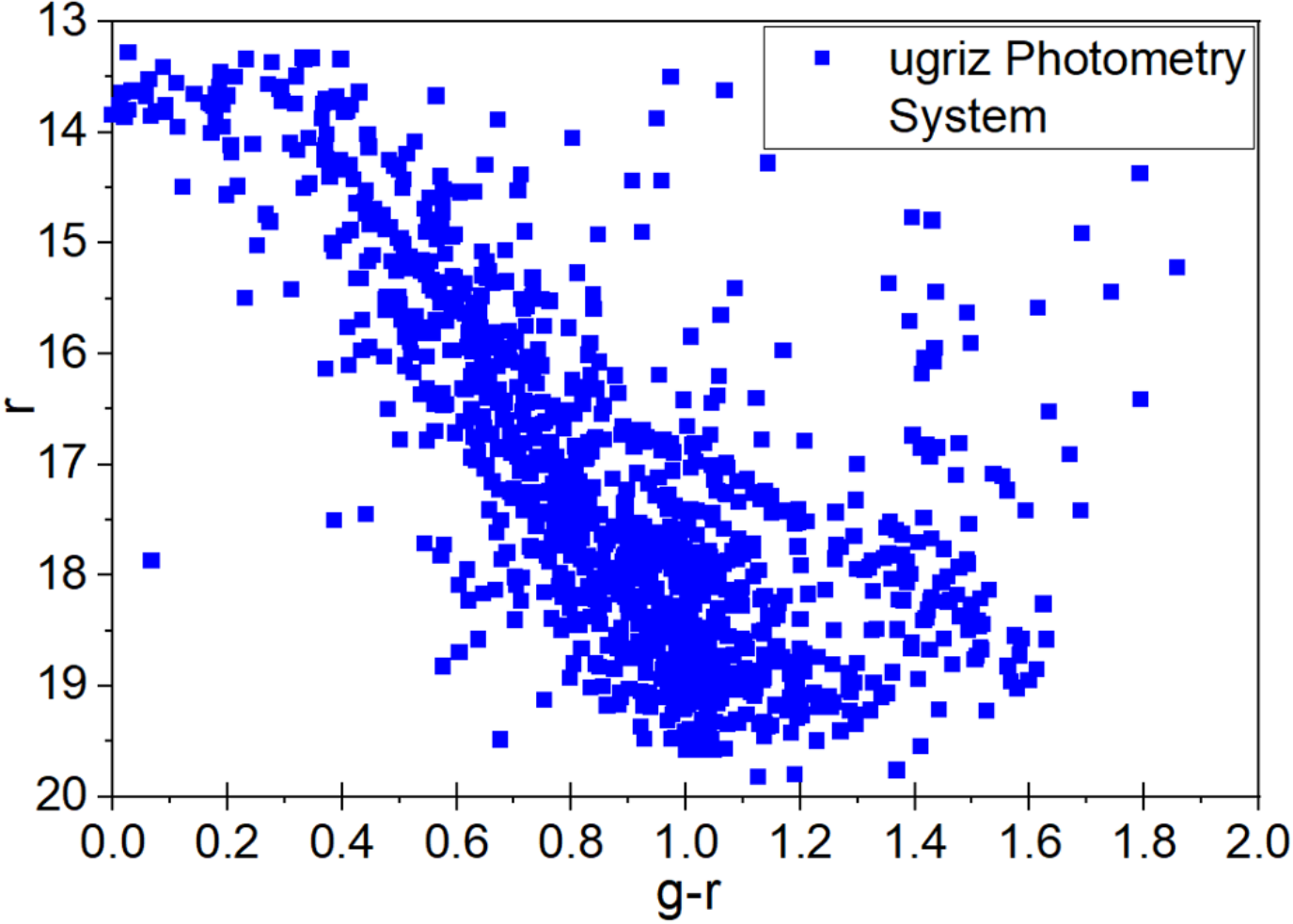}} \label{f2a}}
  \subfigure[Error magnitude plots of \textit{g}, \textit{r}, and \textit{i} magnitudes]
    {\centering{\includegraphics[width=60mm]{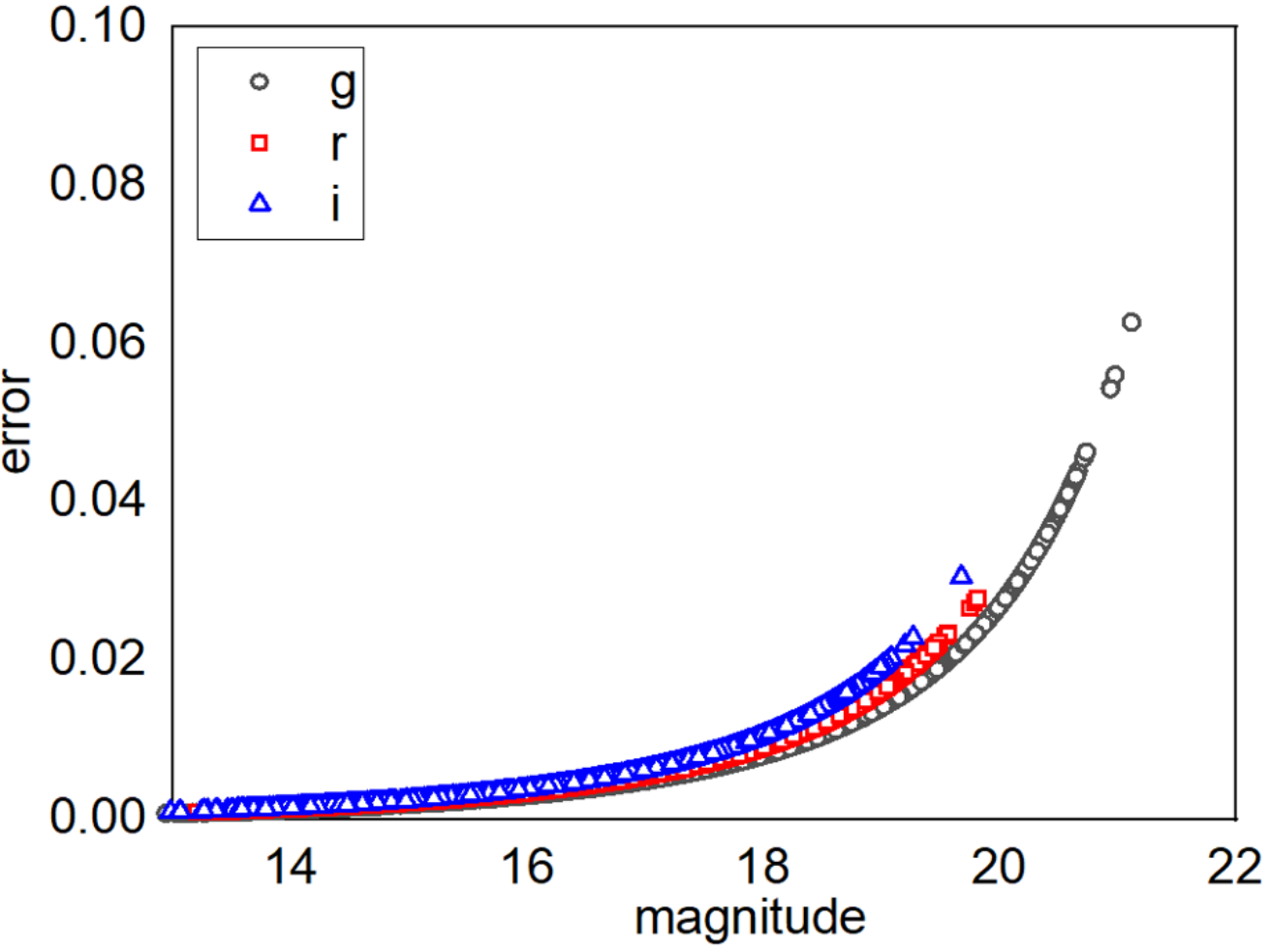}} \label{f2b}}
     \caption[]{Plots of NGC 2360 for ugriz photometric system.} \label{f2}
  \end{center}
\end{figure}

\begin{figure}[!htb]
  \begin{center}
  \subfigure[Color magnitude diagram]
    {\centering{\includegraphics[width=60mm]{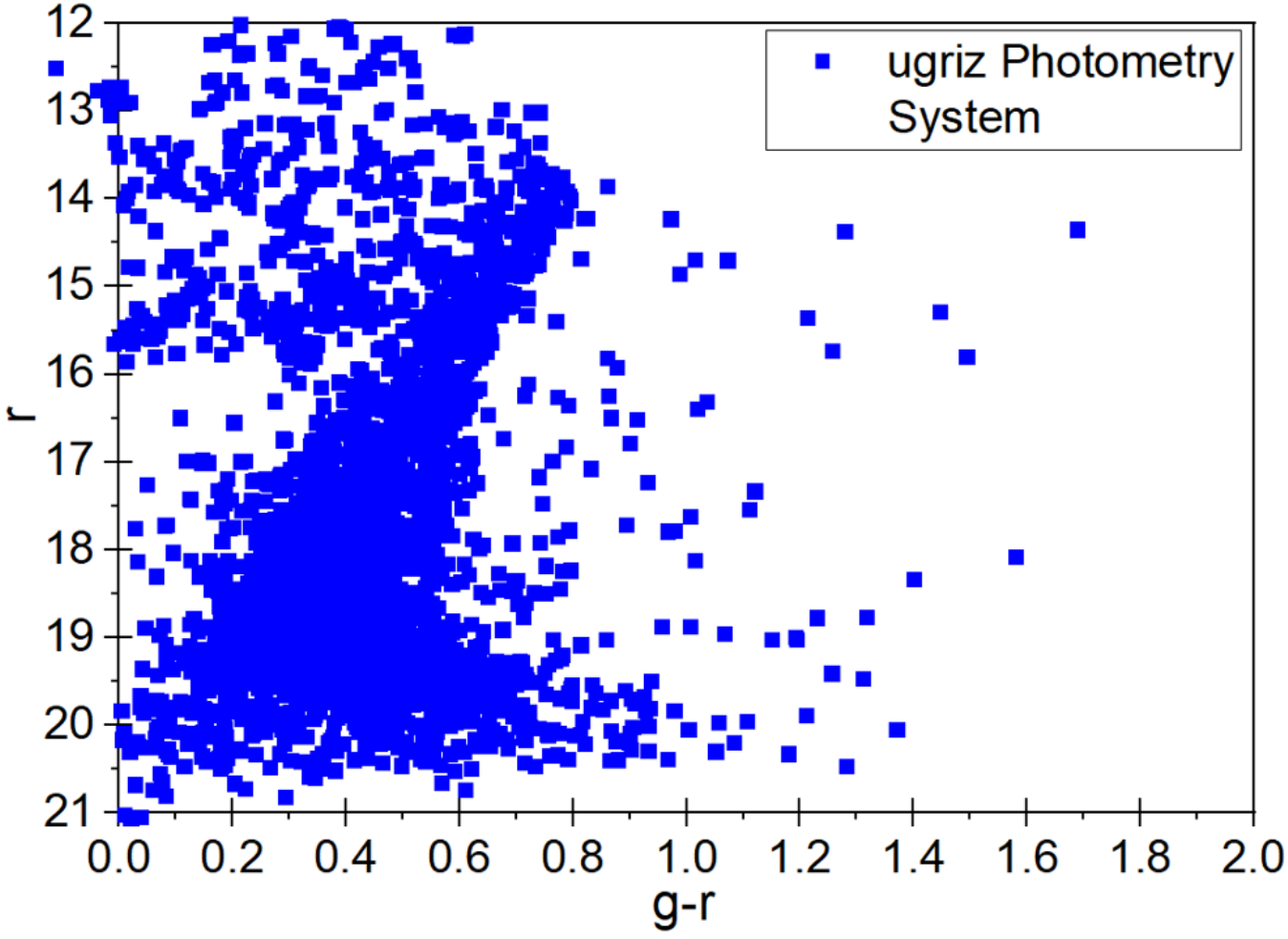}} \label{f3a}}
  \subfigure[Error magnitude plots of \textit{g}, \textit{r}, and \textit{i} magnitudes]
    {\centering{\includegraphics[width=60mm]{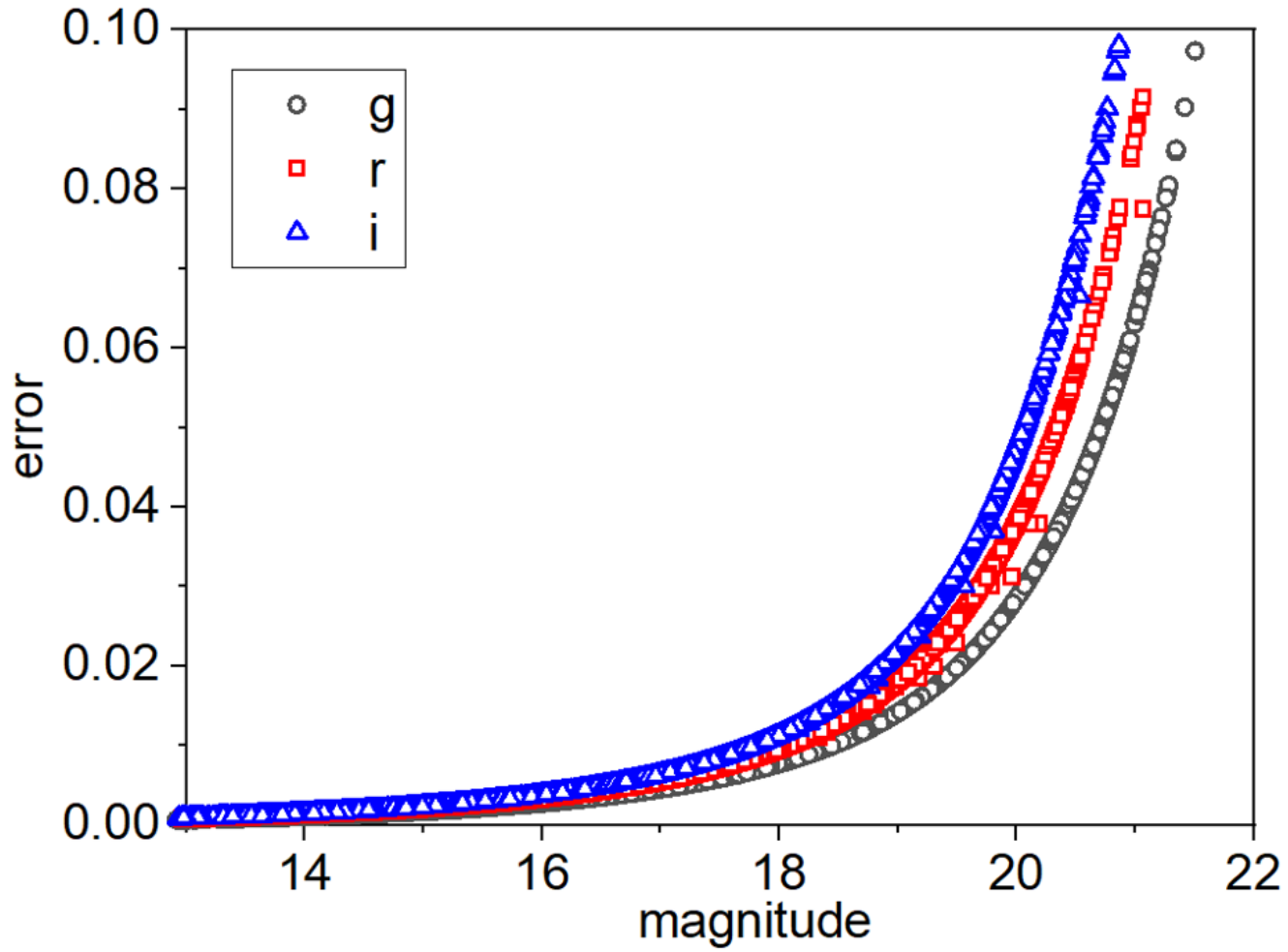}} \label{f3b}}
     \caption[]{Plots of NGC 5272 for ugriz photometric system.} \label{f3}
  \end{center}
\end{figure}

\begin{figure}[!htb]
  \begin{center}
  \subfigure[NGC 2360]
    {\centering{\includegraphics[width=60mm]{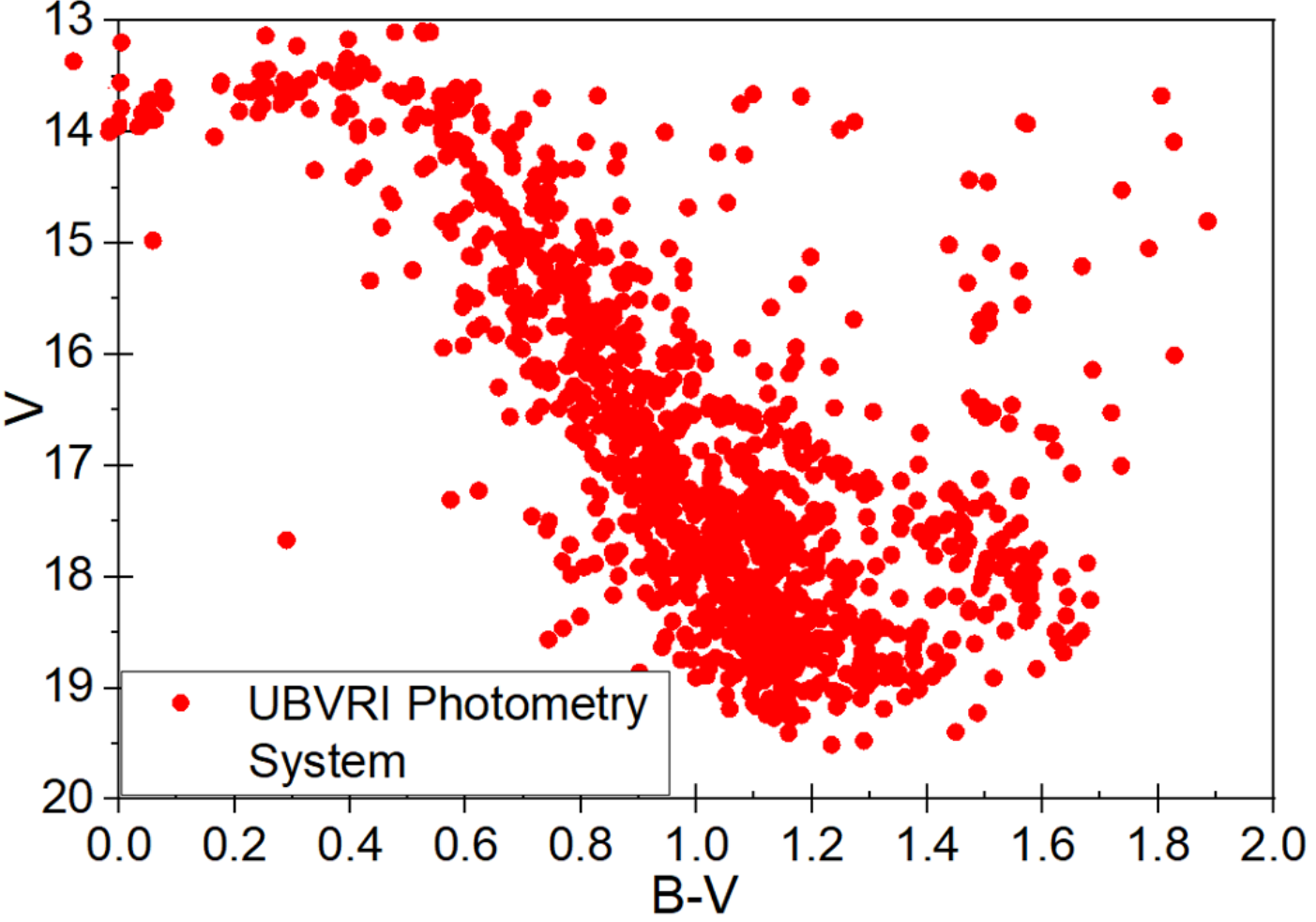}} \label{f4a}}
  \subfigure[NGC 5272]
    {\centering{\includegraphics[width=60mm]{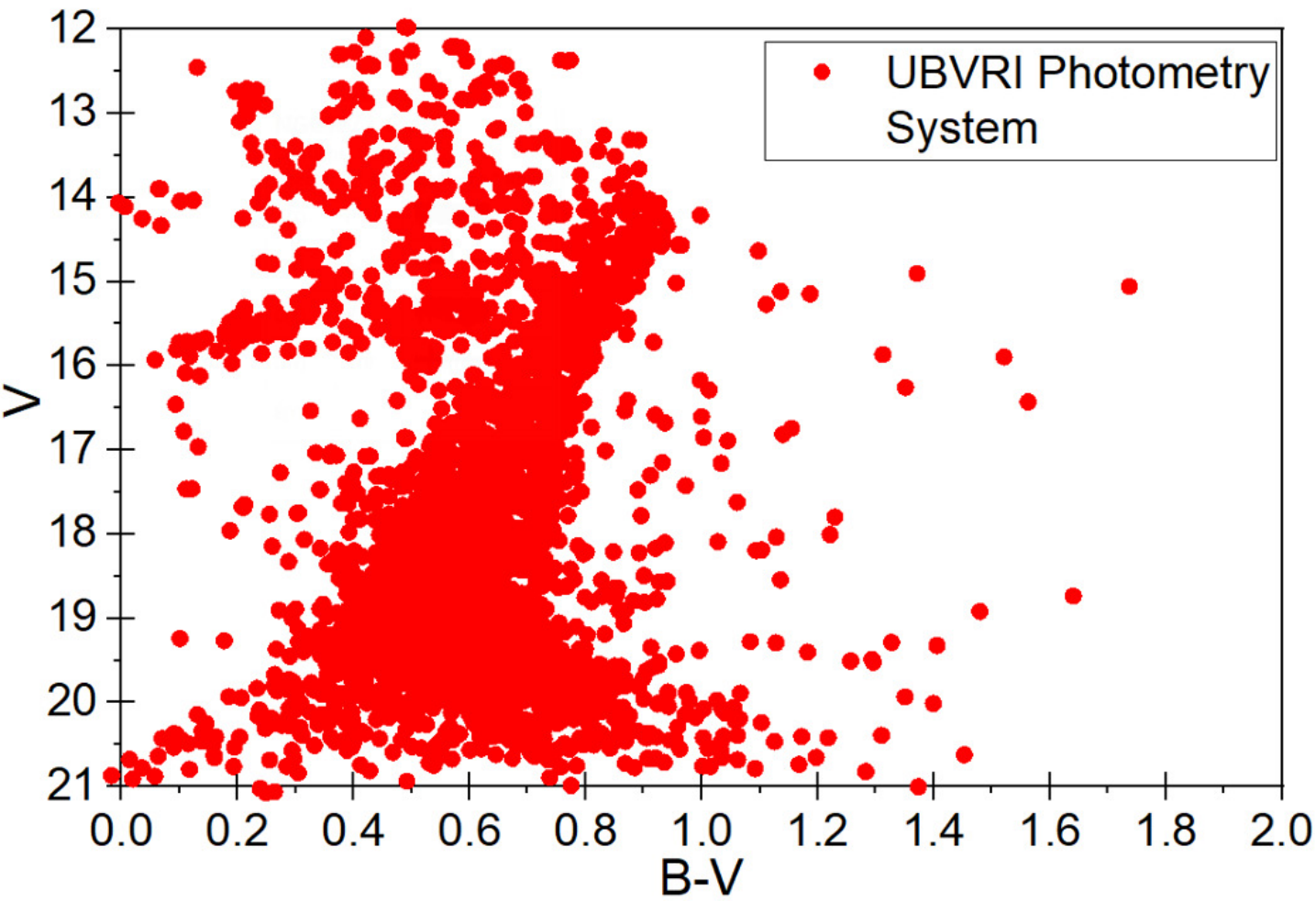}}  \label{f4b}}
     \caption[]{Color magnitude diagram of the selected star clusters for UBVRI photometric system.} \label{f4}
  \end{center}
\end{figure}
}

\section*{4. Stellar Parameters Estimation}
\label{sec:SPE}

{The stellar parameters are evaluated from the CMDs of the chosen star clusters, attained using UBVRI photometry system magnitudes, by fitting the theoretical isochrones on them. Usually, isochrone is defined with absolute magnitudes for different ages, metallicities, and interstellar absorptions in addition to the reddening effects. In order to achieve the essential stellar parameters of the selected star clusters, their respective CMDs are matched with the isochrones of different ages and metallicities until the best fit isochrones are attained. Generally, several isochrones are selected on the basis of previous data (from literature, refer Table \ref{t2}). Then for each model some goodness-of-fit test, for instance, chi-squared test, is performed to obtain the standard deviation, $\sigma$. The essential stellar parameters are then obtained for best-fit values lying within, let's say, $1\sigma$ with lower and upper errors calculated as nth and mth percentile, respectively. The lower and upper errors are required to determine the uncertainties in the stellar parameters. The large numbers of isochrones  would make the analysis computationally heavy. Hence, the manual fitting would not be an ideal method to perform. However, in this work, to comprehend the basic technique behind the estimation of essential parameters, only those isochrones are selected for parameter evaluation that fits the data (CMD) by eye.

With the aim of providing an in-depth procedure to manually match the isochrone with the CMD of a star cluster, different isochrones are matched with NGC 2360 CMD by using TOPCAT software and the fitting is achieved on the basis of visual perception. Also, before finding the best-fit isochrone, CMDs are corrected for the galactic extinctions which are acquired from NASA/IPAC Extragalactic Database. The isochrones utilized in this work are extracted from \url{http://stev.oapd.inaf.it/cgi-bin/cmd_3.4} \cite{Bressan}. To depict the fitting of isochrones, Fig.\,\ref{f5} presents the plotting of isochrones of various ages and metallicities on the CMD of NGC 2360. In Fig.\,\ref{f5a}, isochrones of different ages are obtained for fixed metallicity, [Fe/H], of -0.15, while in Fig.\,\ref{f5b} isochrones of different metallicities for constant log(age) of 8.85 or age of 708 Myrs are acquired. In the CMD of NGC 2360, the isochrones are fitted corresponding to their turn-off point (TO). Eventually, Fig.\,\ref{f6a} depicts the isochrone which is best fitted on the NGC 2360 CMD. This isochrone is having age of 708 Myrs with metallicity, [Fe/H], of -0.15. Successively, aforesaid steps are followed to achieve the best-fit isochrone for NGC 5272. Specifically, in the case of NGC 5272, the horizontal branch (HB) is considered for obtaining the best-fit isochrone. The optimum isochrone for NGC 5272 is exhibited in Fig.\,\ref{f6b}. It is observed that the best-fit isochrone for NGC 5272 is one which is having log(age) of 10.063 or age of 11.56 Gyrs with metallicity, [Fe/H], of -1.57. 

\begin{figure}[!htb]
  \begin{center}
  \subfigure[Isochrones with ${[Fe/H]}$=-0.15 and different ages]
    {\centering{\includegraphics[width=60mm]{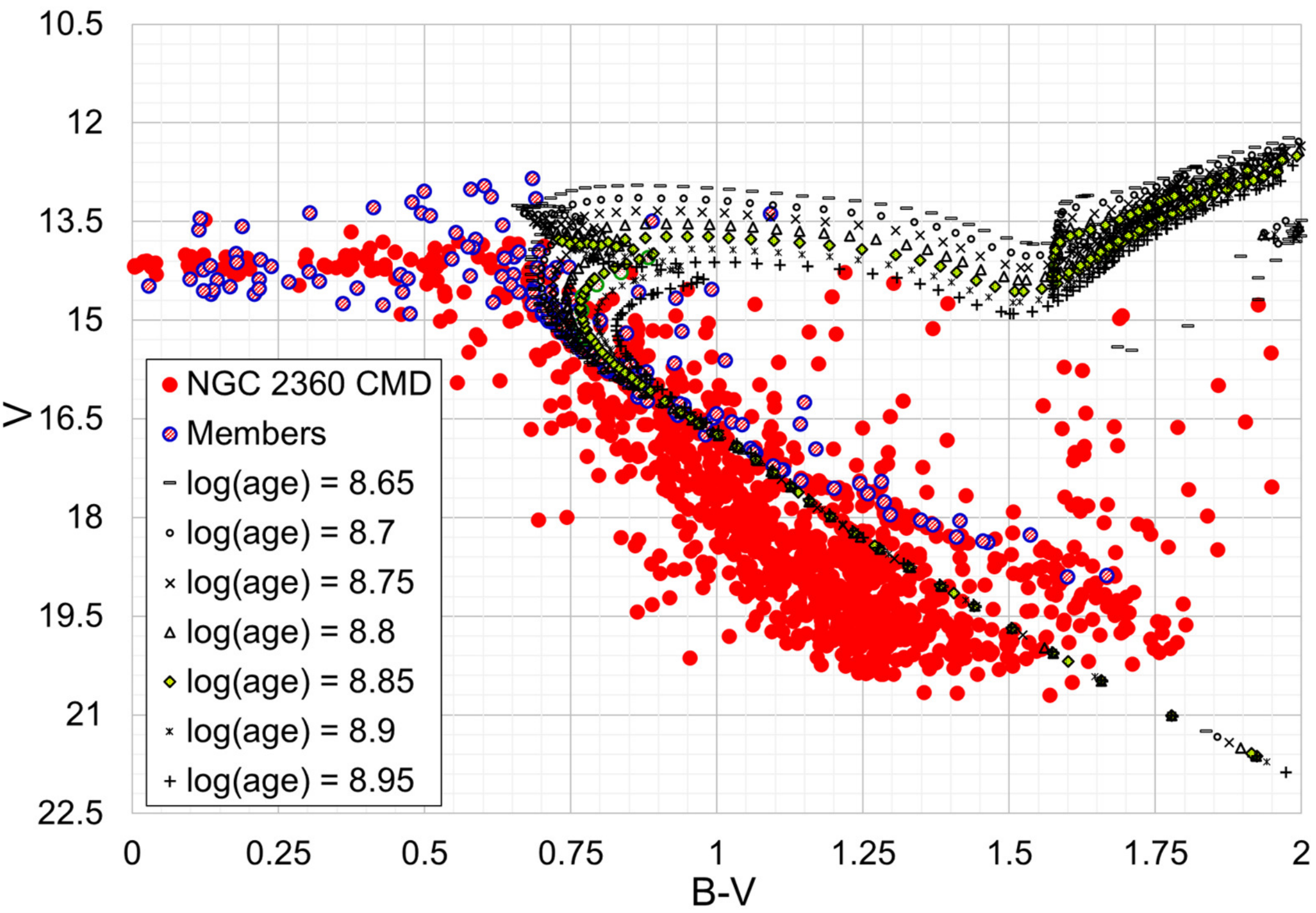}} \label{f5a}}
  \subfigure[Isochrones with different ${[Fe/H]}$ and log(age) = 8.85 or age = 708 Myrs]
    {\centering{\includegraphics[width=60mm]{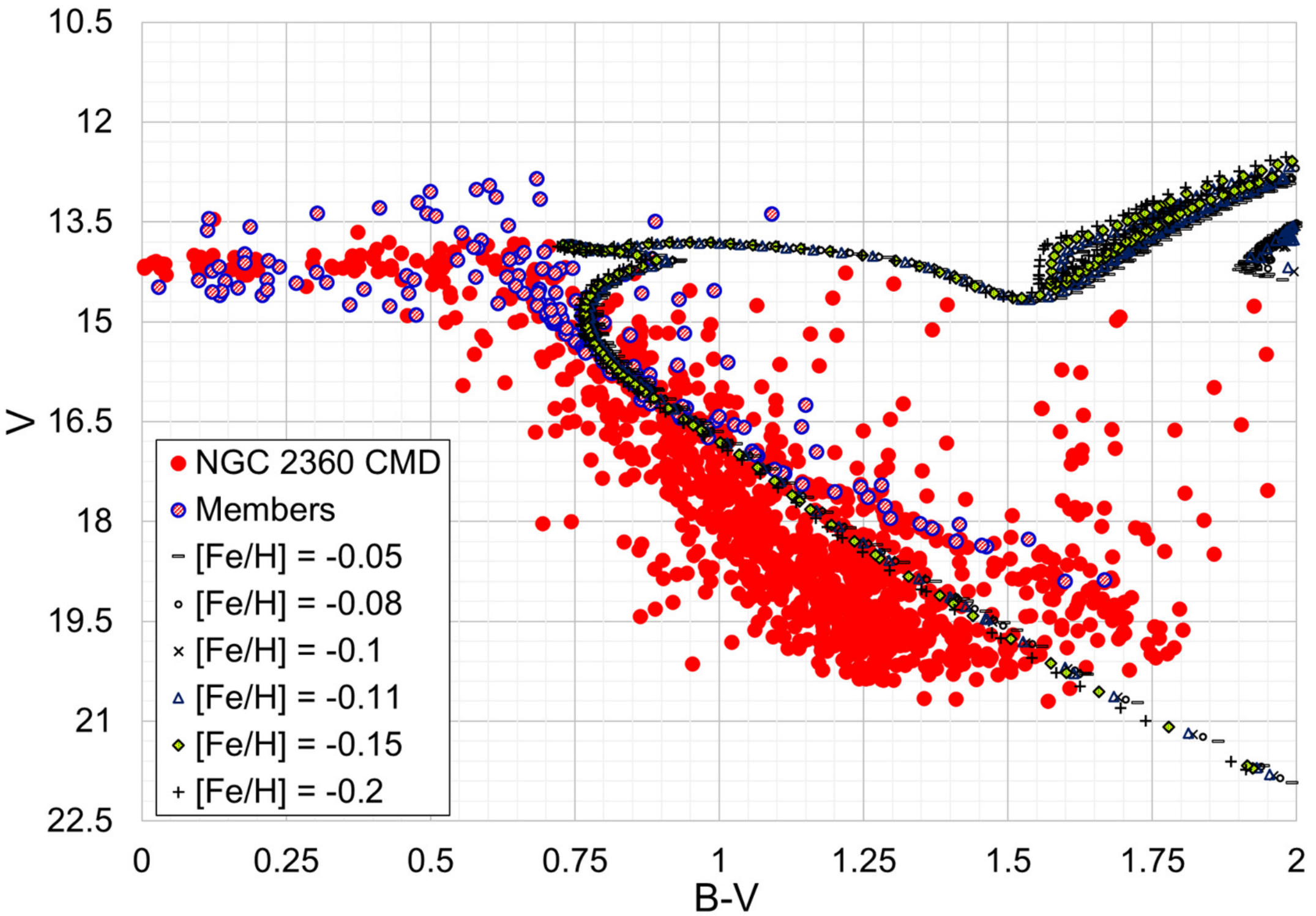}}  \label{f5b}}
     \caption[]{Fitting of isochrones with different metallicities and different ages on the $B-V$ vs $V$ CMD of NGC 2360.} \label{f5}
  \end{center}
\end{figure}

\begin{figure}[!htb]
  \begin{center}
  \subfigure[NGC 2360, ${[Fe/H]}$=-0.15]
    {\centering{\includegraphics[width=60mm]{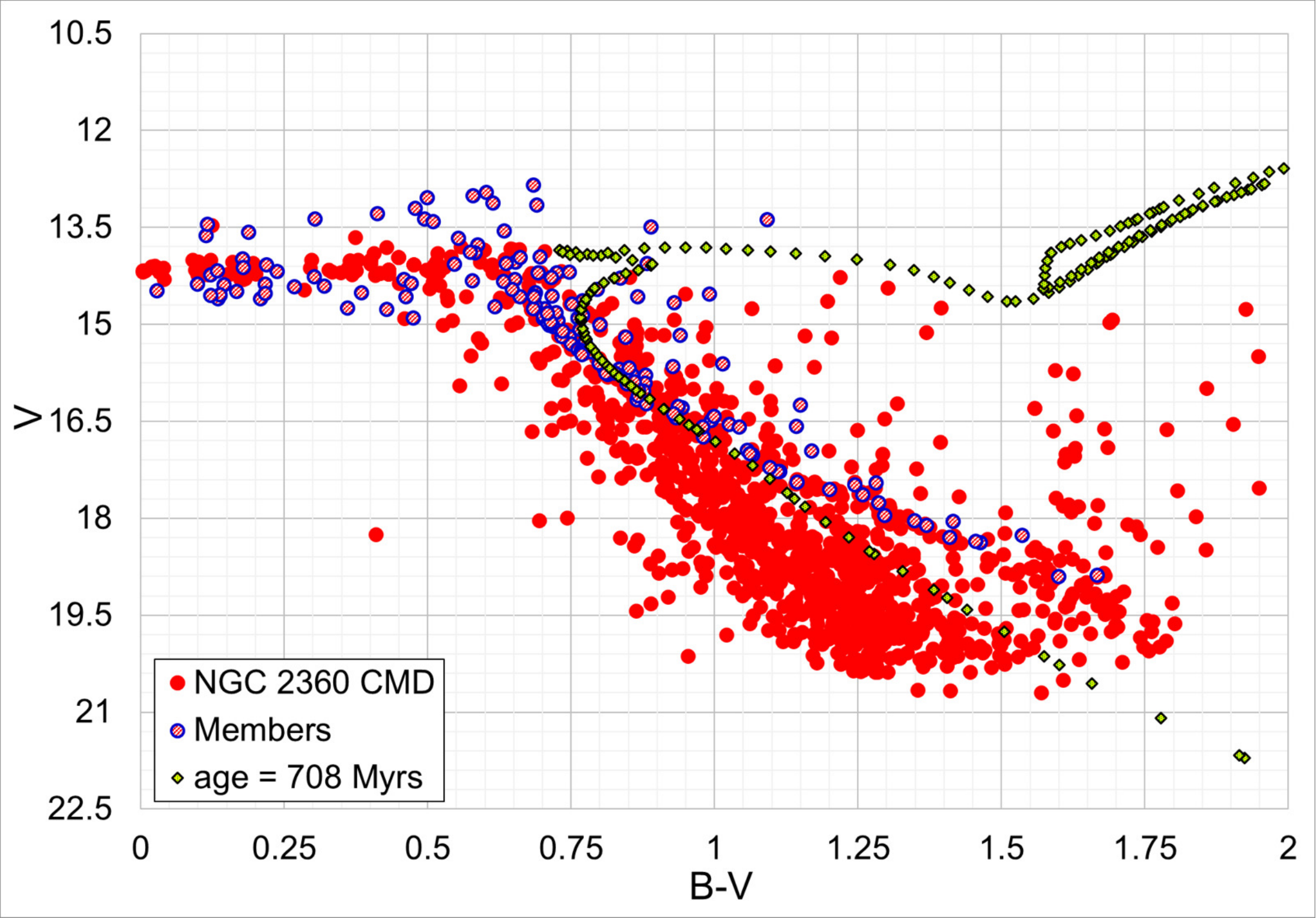}} \label{f6a}}
  \subfigure[NGC 5272, ${[Fe/H]}$=-1.57]
    {\centering{\includegraphics[width=60mm]{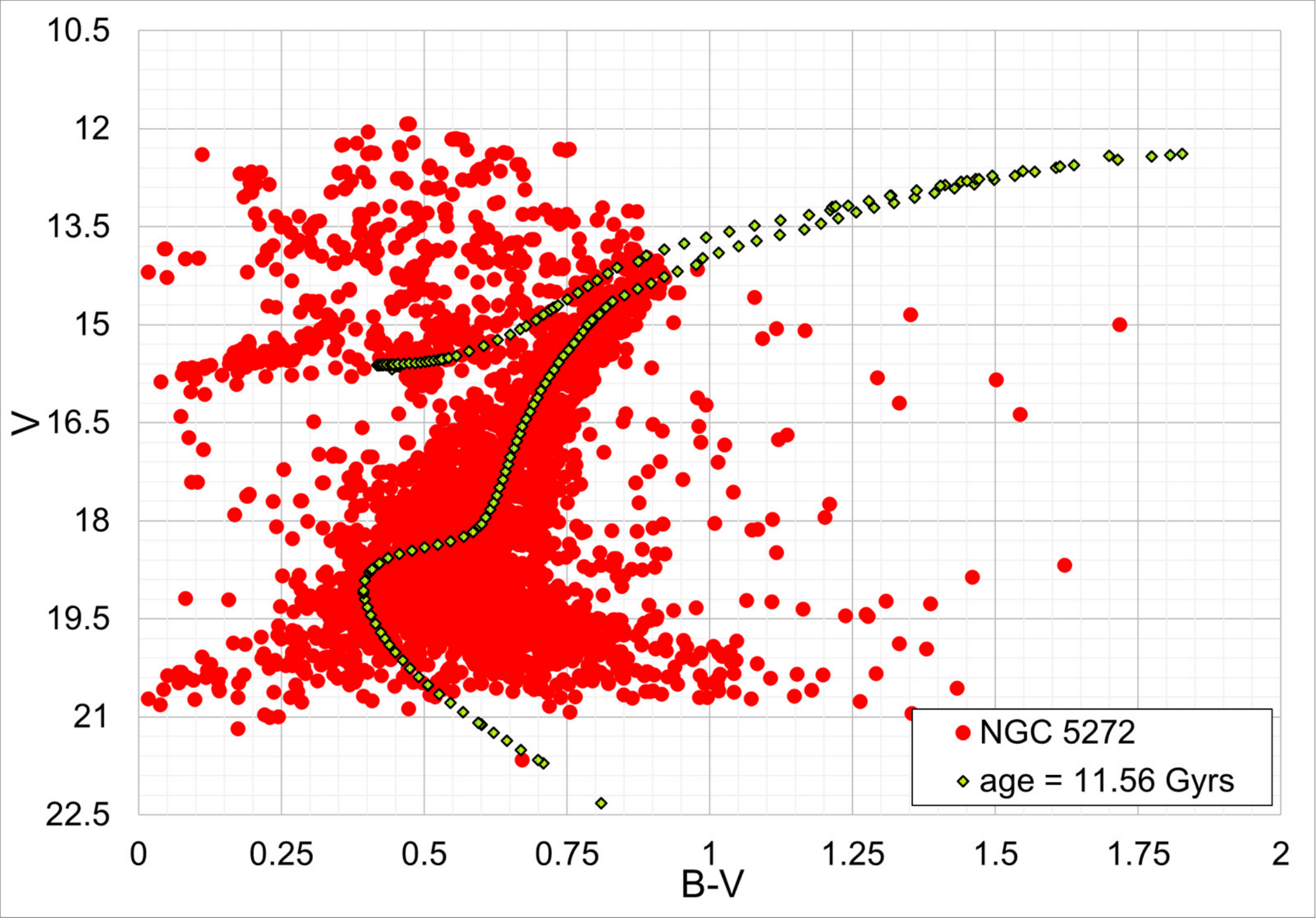}} \label{f6b}}
     \caption[]{Optimized isochrone fitted on the respective CMDs of NGC 2360 and NGC 5272.} \label{f6}
  \end{center}
\end{figure}

Furthermore, it should be noted that the shift in color indices ($B-V$), performed to obtain the optimum isochrone, provides the reddening parameter value, whereas the shift in the apparent magnitudes ($V$) gives the value of distance modulus. While fitting the isochrones, interstellar reddening, $E(B-V)$, and distance modulus, $DM$, for NGC 2360 are obtained as 0.12 and 11.65, respectively. Whereas, for NGC 5272, the interstellar reddening is determined as $E(B-V)$=0.015, and the distance modulus is found to be $DM$=15.1. In addition, the interstellar extinction absorption, $A(V)$, in the visual band and the distance, $d$, of the selected star clusters are evaluated from the interstellar reddening and distance modulus values. The determined values of $A(V)$ and $d$ for NGC 2360 are 0.372 and 1.801, respectively, while for NGC 5272 the values are calculated as $A(V)$=0.0465 and $d$=10.249. The evaluated stellar parameters of NGC 2360 and NGC 5272 from their CMDs are summarized in Table \ref{t2}. It is observed from Table \ref{t2} that the values of these stellar parameters are in close approximation with the results reported in the literature which are based on the IRAF analysis technique.

\begin{table}[htbp]
\small
 \begin{center}
  \caption{Stellar parameters values for NGC 2360 and NGC 5272 compared with the results reported in literature.}
  \label{t2}
  \resizebox{\textwidth}{!}{
   \begin{tabular}{lcccccc}
    \hline 
    \multicolumn{1}{l}{\multirow{2}[0]{*}{\textbf{Stellar}}} & \multicolumn{1}{c}{\multirow{2}[0]{*}{\textbf{NGC 2360}}} & \multicolumn{3}{c}{\textbf{NGC 5272}} & \multicolumn{2}{c}{\textbf{Star Clusters Studied}} \\
     \cline{3-5} \cline{6-7} &        &
     \multicolumn{1}{c}{\textbf{Harris,}} & \multicolumn{2}{c}{\textbf{Dodd et al.,}} & \textbf{NGC} & \textbf{NGC} \\
      \textbf{Parameters} &  (Webda)    & 1996  & \multicolumn{2}{c}{2019}   & 2360 & 5272   \\
          \hline
    \textbf{t (Gyr)} & –     & –     & 9.5   & 11    & 0.7079 & 11.56 \\
    \textbf{log(age)} & 8.749 & –     & –     & –     & 8.85   & 10.063 \\
    \textbf{[Fe/H]} & -0.15 & -1.5  & -2.5  & -0.5  & -0.15 & -1.57 \\
    \textbf{Distance Modulus} & 11.72 & 15.07 & –     & –     & 11.65 & 15.1 \\
    \textbf{E(B-V) (mag)} & 0.111 & 0.01  & 0     & 0     & 0.12 & 0.015 \\
    \textbf{A(V)} & –     & –     & –     & –     & 0.372 & 0.0465 \\
    \textbf{Distance (kpc)} & 1.887 & 10.2  & 13    & 5     & 1.801 & 10.249 \\
    \hline
   \end{tabular}}
      \end{center}
  \end{table}

Along with the above-mentioned stellar parameters, the graphs depicting the distributions of some other parameters are also presented subsequently. For plotting the graphs of various other parameter distributions, initially, the CMDs are transformed into the Hertzsprung–Russell diagram (HR diagram). HR diagram or $T_{eff}-L$ diagram is one in which colors are related to effective temperatures ($T_{eff}$) and the stars' magnitudes with known distance are related to the luminosities ($L$). For plotting the HR diagram, the effective temperature of the hotter and cooler stars presented in the selected star clusters are evaluated using equations: Eq. (\ref{r5}) and Eq.(\ref{r6}) \cite{Sekiguchi, Gray, Ballesteros}. The luminosity is obtained by first finding the linear radius of the stars from the relationship represented by Eq. (\ref{r7}) \cite{Lamers} with $L_{\bigodot}$=$3.85 \times 10^{33} \: erg \: s^{-1}$ and $R_{\bigodot}$=$6.9598 \times 10^{8} \: m$. On the other hand, the linear radius $R$ is also related to the angular radius, $\theta_R$, (Refer Eq. \ref{r9}) which in turn is related to the surface brightness, $\mathbb{P}_{V}$, and the apparent magnitude, $m_V$, with ${\theta_R}^{\bigodot}=959.63 ~arcsec$ and $m_{V}^{\bigodot}=-26.75$ (Refer Eq. \ref{r10}). The equations: Eq. (\ref{r8}) through Eq. (\ref{r11}) adopted from \cite{Gray, Seeds, Lamers} are utilized to obtain linear radius, angular radius, and surface brightness of the sources present in the selected star clusters. The linear radius and the angular radius relationship could be rewritten in terms of the solar parameter as presented in Eq. (\ref{r9}). Further, the mass of the stars in selected star clusters could be achieved from either Eq. (\ref{r12}) or from Eq. (\ref{r13}) by substituting the luminosity or linear radius values, respectively \cite{Neece}. 

\begin{eqnarray}
 \label{r5}
 \begin{aligned}
    log(T_{eff}) & = 3.981+0.0142(B-V)+16.3618(B-V)^2 \\  
    & +81.891(B-V)^3+161.5075(B-V)^4 \\ & \: for \: (B-V)<0.00,
    \end{aligned}
\end{eqnarray}

\begin{eqnarray}
 \label{r6}
    \begin{aligned}
    log(T_{eff}) & = 3.981-0.4728(B-V)+0.2434(B-V)^2 \\ &
    -0.0620(B-V)^3 \: for \: 0.0 \leq (B-V)\geq1.5,
    \end{aligned}
\end{eqnarray}

\begin{eqnarray}
 \label{r7}
 \begin{aligned}
    \frac{L}{L_{\bigodot}} &= \left(\frac{R}{R_{\bigodot}}\right)^2 \left(\frac{T_{eff}}{5777K}\right)^4,
    \end{aligned}
\end{eqnarray}

\begin{eqnarray}
 \label{r8}
 \begin{aligned}
    \frac{angular \: radius}{206265} &= \left(\frac{linear \: radius}{distance}\right),
    \end{aligned}
\end{eqnarray}

\begin{eqnarray}
 \label{r9}
 \begin{aligned}
    \frac{R}{R_{\bigodot}} &= \left(\frac{\theta_R}{206265}\right) \left(\frac{d}{R_{\bigodot}}\right),
    \end{aligned}
\end{eqnarray}

\begin{eqnarray}
 \label{r10}
 \begin{aligned}
    \mathbb{P}_{V} &= -0.1(m_{V}-m_{V}^{\bigodot})-0.5log\left(\frac{\theta_R}{{\theta_R}^{\bigodot}}\right),
    \end{aligned}
\end{eqnarray}

\begin{eqnarray}
 \label{r11}
 \begin{aligned}
    \mathbb{P}_{V} & = 0.2241-0.5610(B-V)+0.6207(B-V)^2 \\  
    & -0.6056(B-V)^3+0.2041(B-V)^4,
    \end{aligned}
\end{eqnarray}

\begin{eqnarray}
 \label{r12}
    \begin{aligned}
    \frac{L}{L_{\bigodot}} & \approx & 0.231\left(\frac{M}{M_{\bigodot}}\right)^{2.61},
    \end{aligned}
\end{eqnarray}

\begin{eqnarray}
 \label{r13}
    \begin{aligned}
    \frac{R}{R_{\bigodot}} & \approx & 0.876\left(\frac{M}{M_{\bigodot}}\right)^{0.807},
    \end{aligned}
\end{eqnarray}

The effective temperature obtained from equations: Eq. (\ref{r5}) and Eq. (\ref{r6}) for NGC 2360 and NGC 5272 are plotted as a function of $B-V$ color in Figs.\,\ref{f7a} and \,\ref{f7b}, respectively. These graphs of Figs.\,\ref{f7} depict the correlation between the effective temperature relationships defined by equations: Eq. (\ref{r5}) and Eq.(\ref{r6}) for the chosen star clusters. Further, the HR diagrams for NGC 2360 and NGC 5272 are represented in Fig.\,\ref{f8} and it could be observed that these HR diagrams are very similar to the CMDs of the respective clusters depicted in Fig.\,\ref{f4}. Thus, it could be stated that the HR diagram is one with $log (T_{eff})$ on the x-axis and $log (L/L_{\bigodot})$ on the y-axis while CMD has a color on the x-axis and a magnitude (absolute or apparent) on the y-axis. Subsequently, Fig.\,\ref{f9} represents the surface brightness factor, $\mathbb{P}_{V}$, as a function of the color index, $\textit(B-V)$, for all the sources determined by APT software in NGC 2360 and NGC 5272. Furthermore, distributions of the stars' radii, masses, and temperatures for both the star clusters, attained using the surface brightness method, are plotted which are depicted in Fig.\,\ref{f10} and Fig.\,\ref{f11}, respectively.

\begin{figure}[!htb]
  \begin{center}
  \subfigure[NGC 2360]
    {\centering{\includegraphics[width=60mm]{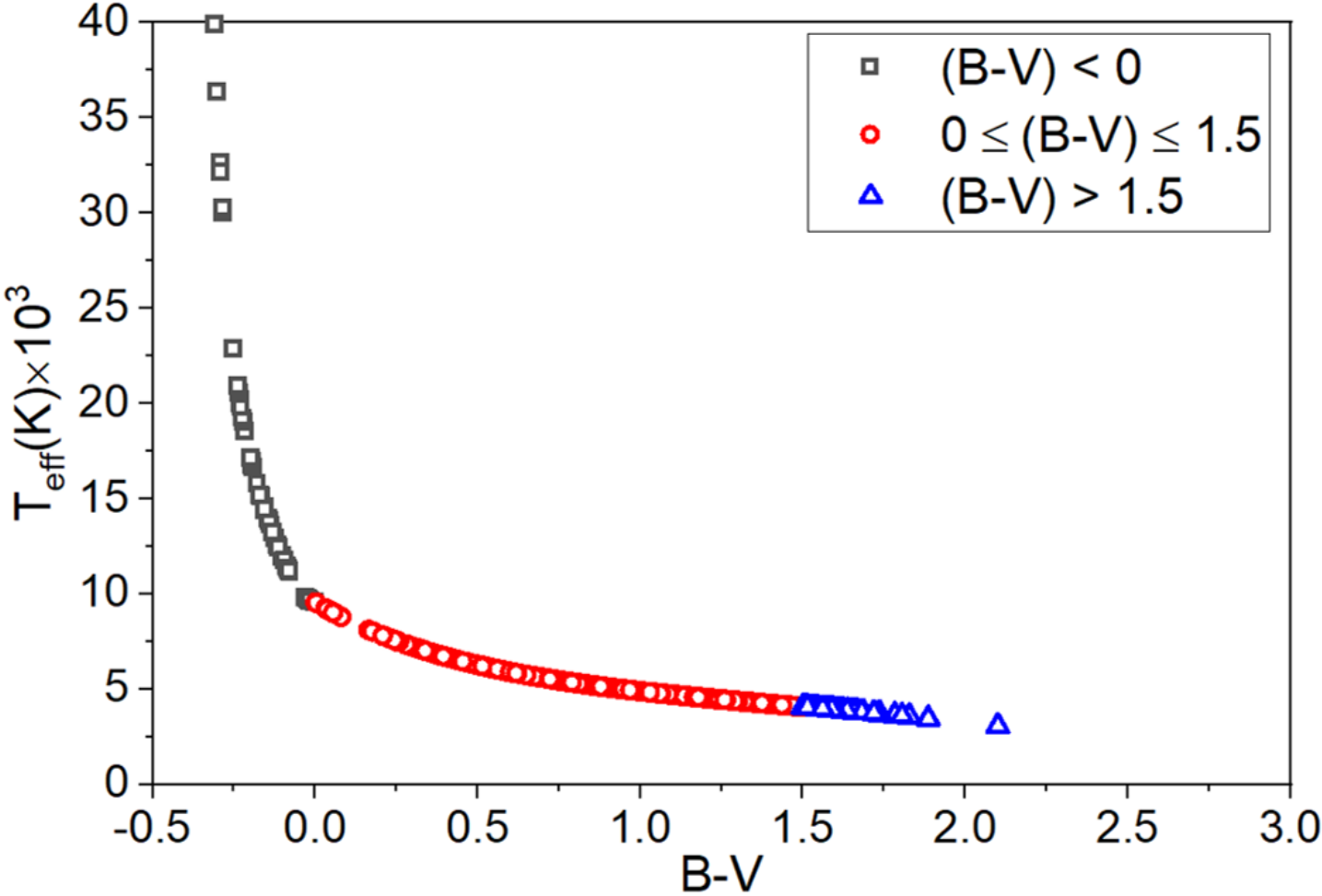}} \label{f7a}}
  \subfigure[NGC 5272]
    {\centering{\includegraphics[width=60mm]{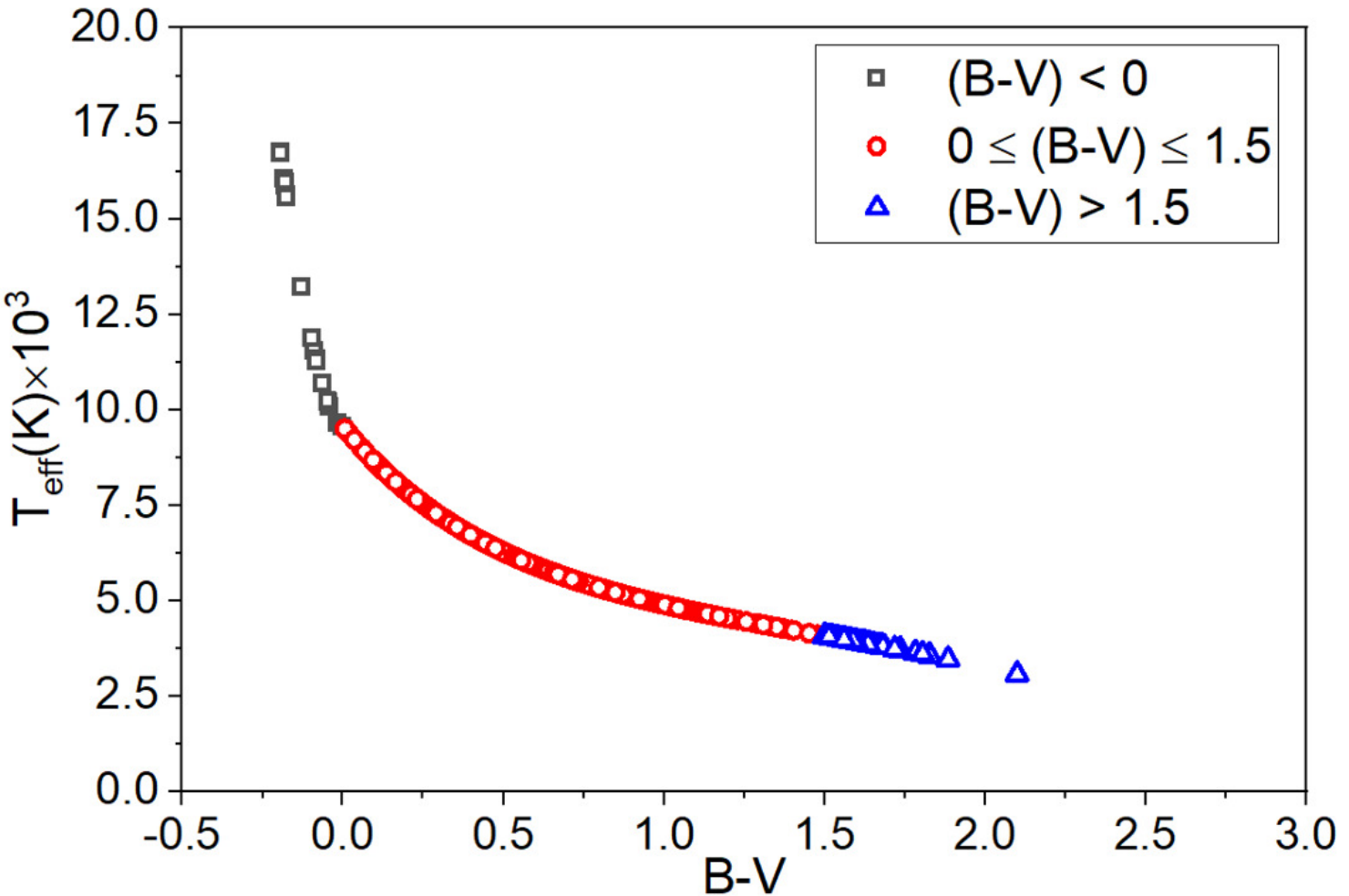}} \label{f7b}}
     \caption[]{The effective temperature as a function of $B-V$ color from equations (\ref{r5}) and (\ref{r6}).} \label{f7}
  \end{center}
\end{figure}

\begin{figure}[!htb]
  \begin{center}
  \subfigure[NGC 2360]
    {\centering{\includegraphics[width=60mm]{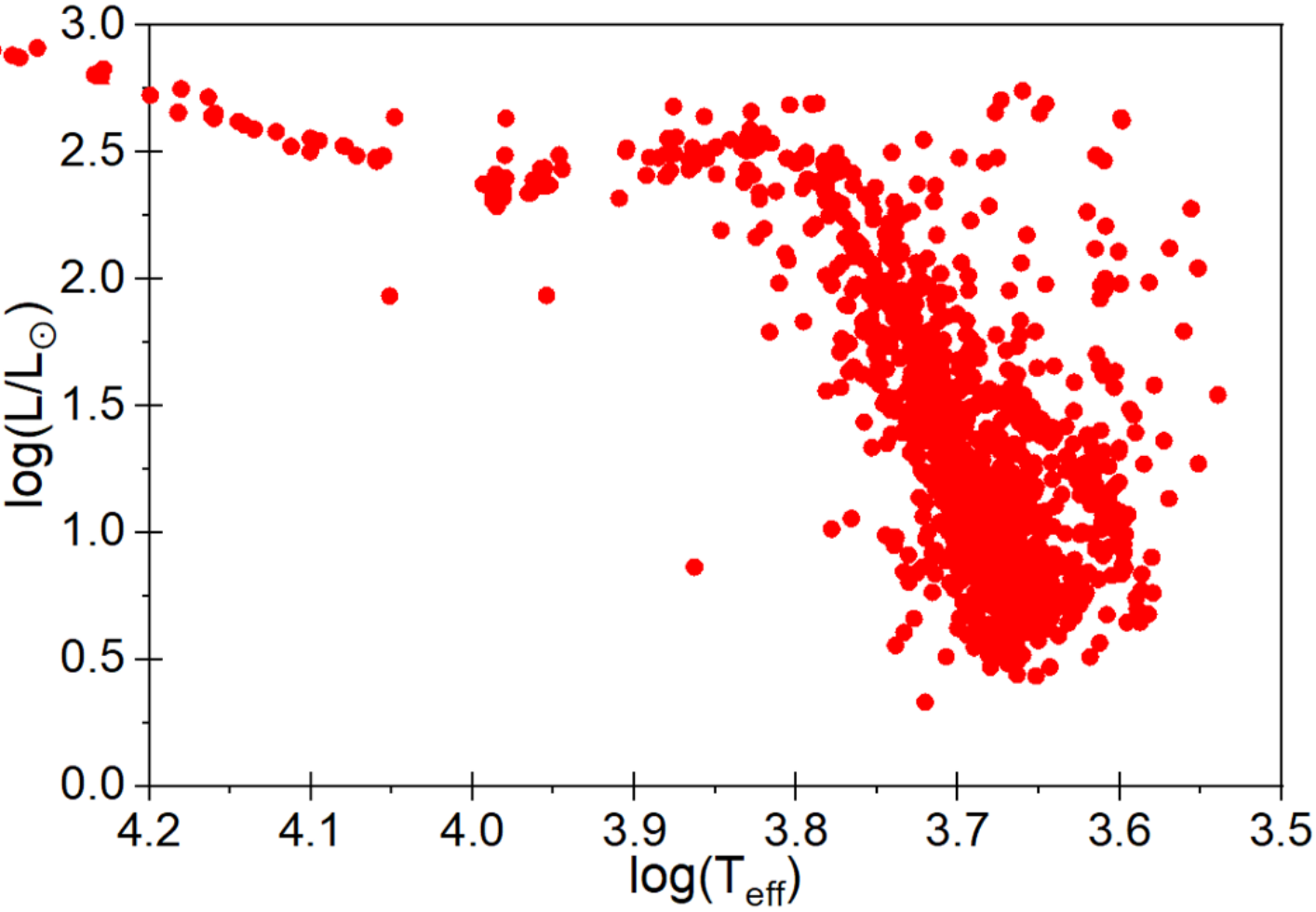}} \label{f8a}}
  \subfigure[NGC 5272]
    {\centering{\includegraphics[width=60mm]{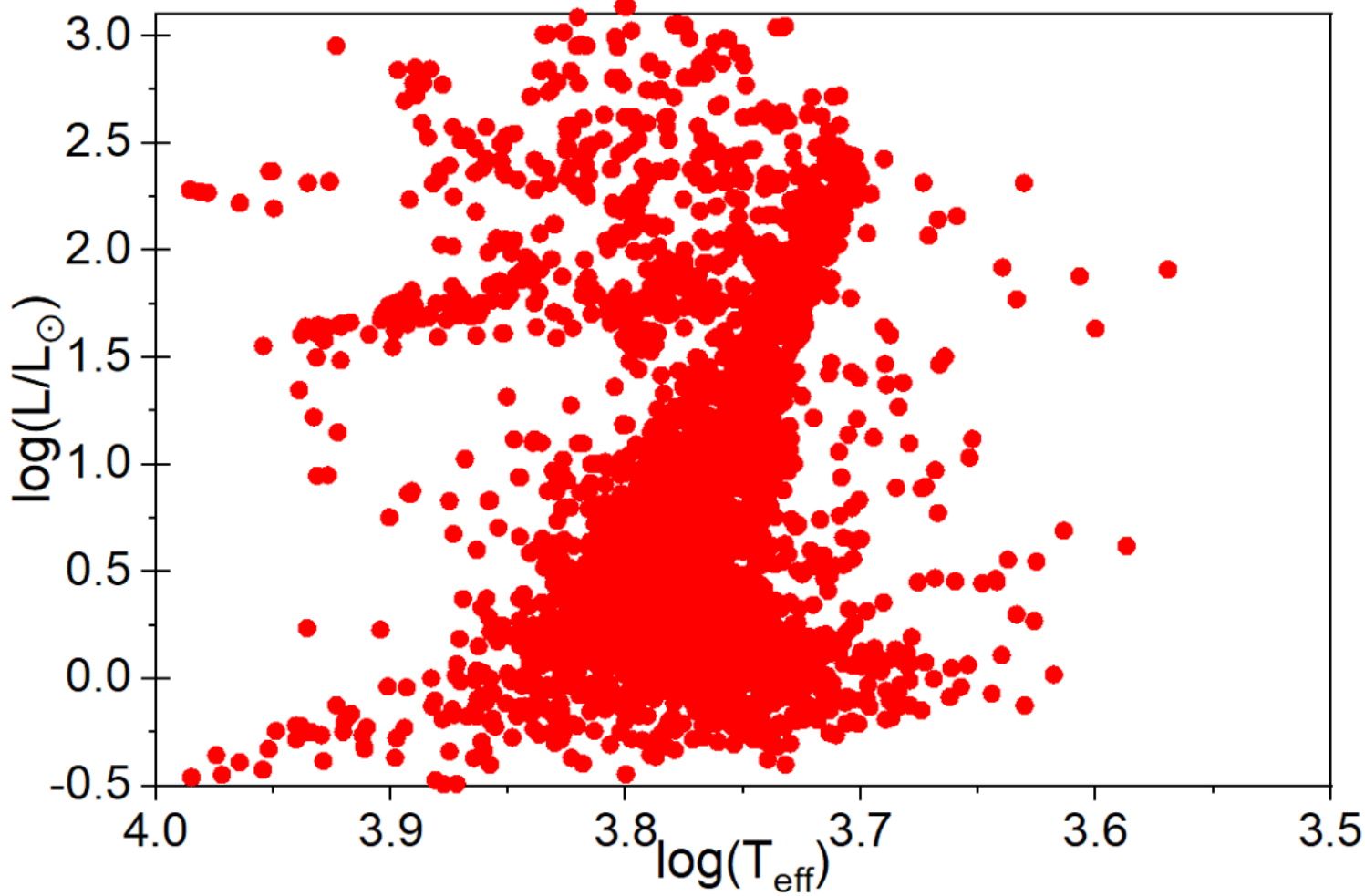}} \label{f8b}}
     \caption[]{HR diagrams of open cluster and globular cluster.} \label{f8}
  \end{center}
\end{figure}

\begin{figure}[!htb]
  \begin{center}
  \subfigure[NGC 2360]
    {\centering{\includegraphics[width=60mm]{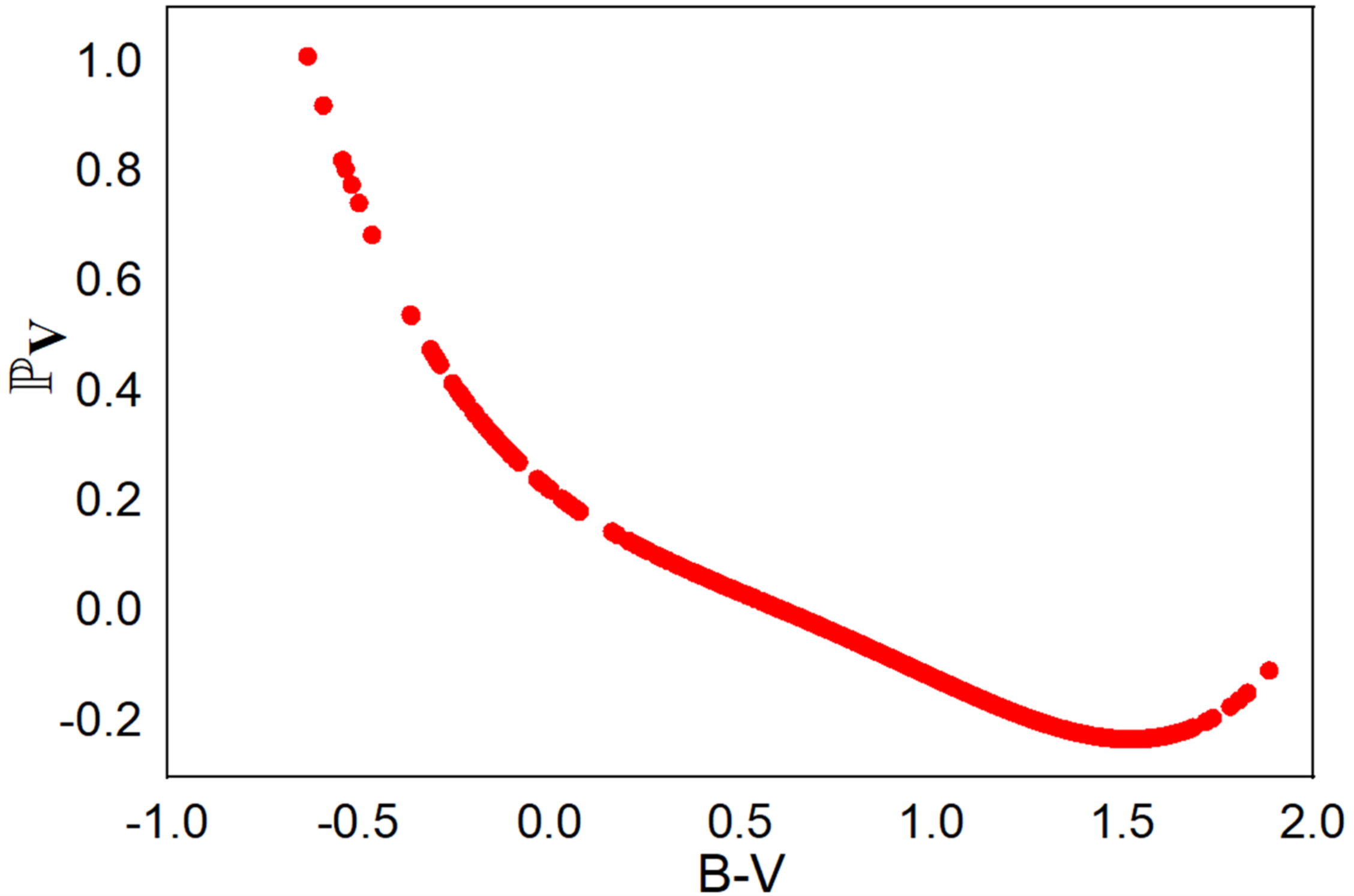}} \label{f9a}}
  \subfigure[NGC 5272]
    {\centering{\includegraphics[width=60mm]{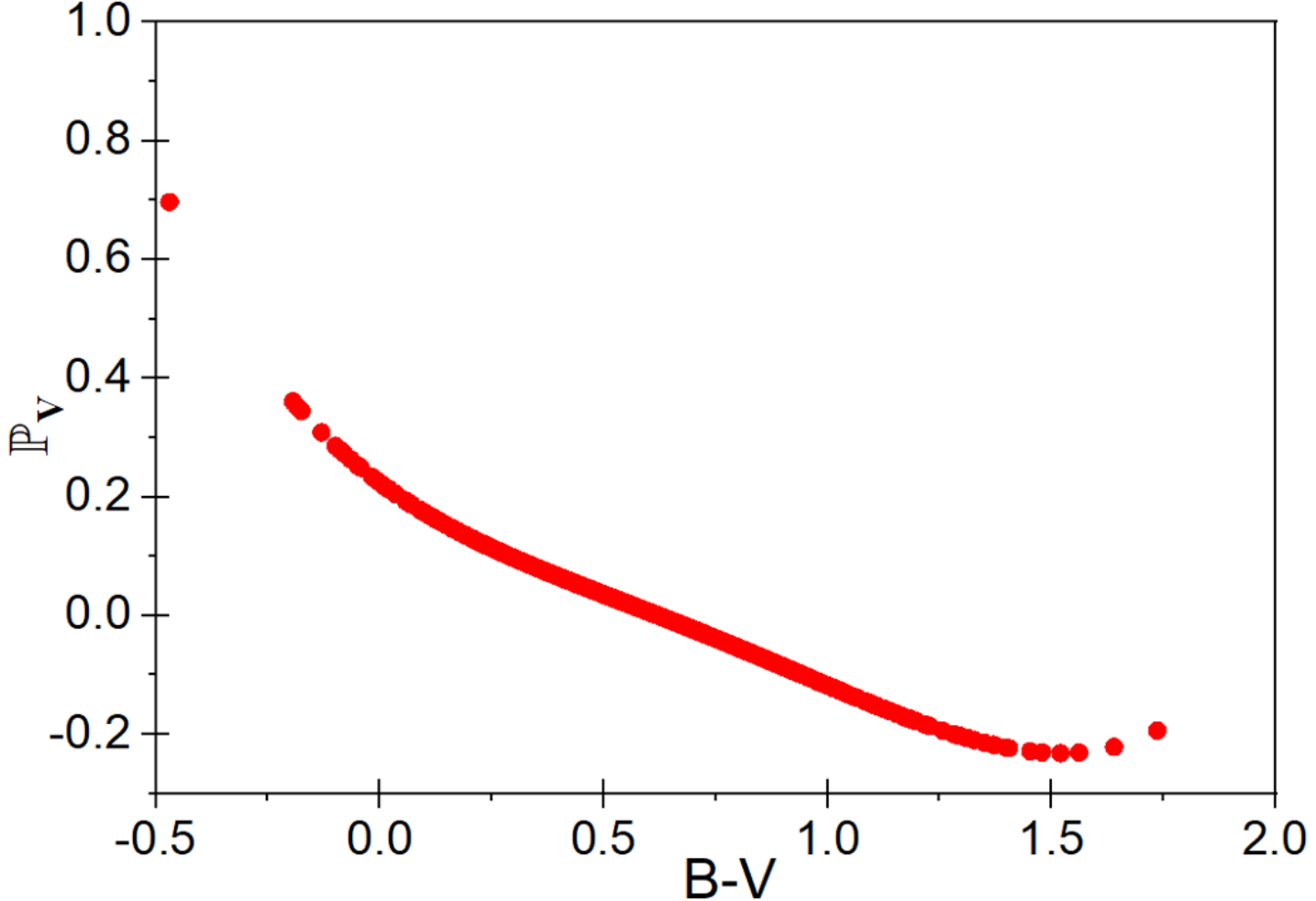}} \label{f9b}}
     \caption[]{The surface brightness factor, $\mathbb{P}_{V}$, as a function of color index $\textit(B-V)$ for NGC 2360 and NGC 5272.} \label{f9}
  \end{center}
\end{figure}

\begin{figure}[!htb]
  \begin{center}
  \subfigure[Distribution of the radii $({R}/{R_{\bigodot}})$]
    {\centering{\includegraphics[width=60mm]{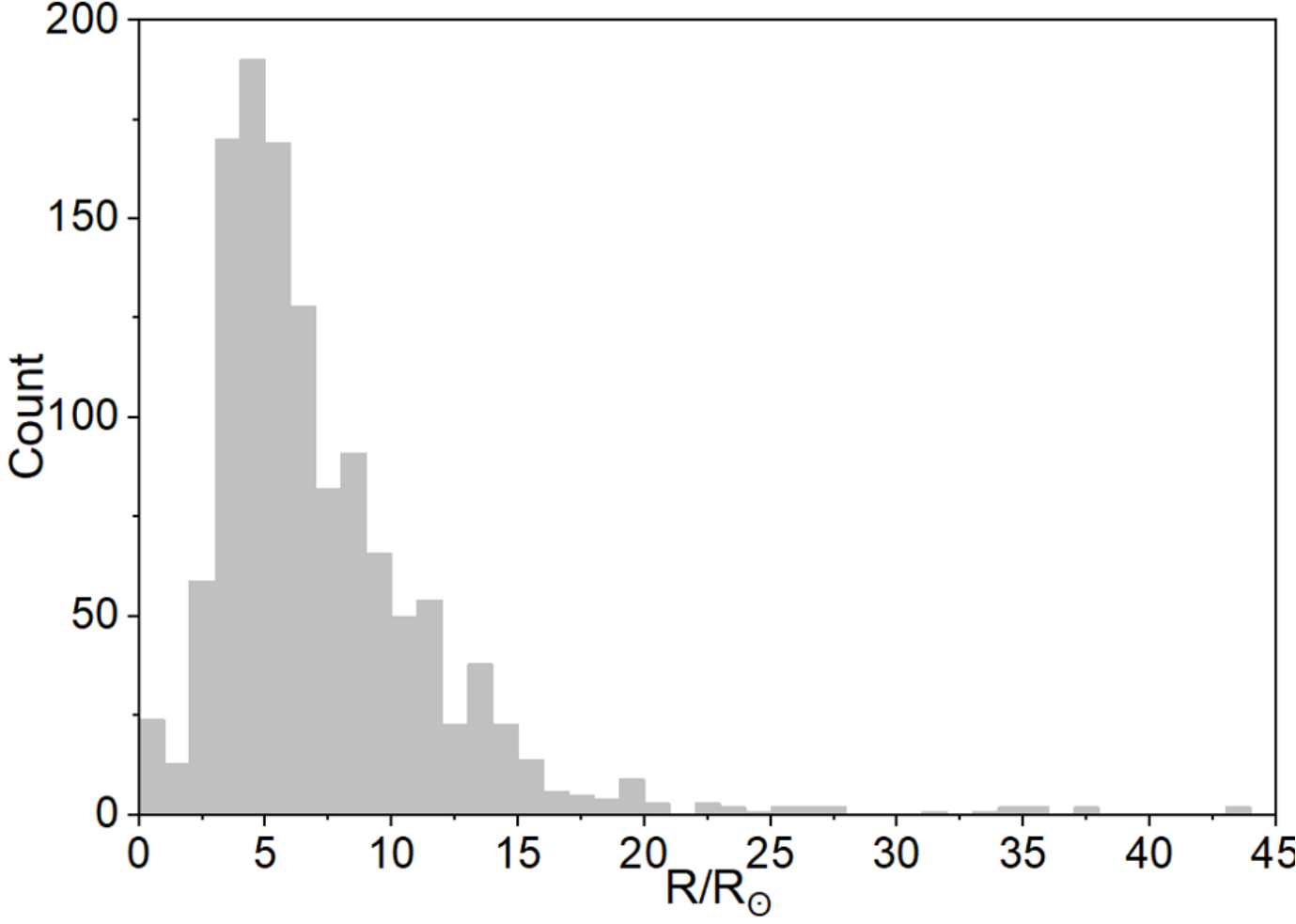}} \label{f10a}}
  \subfigure[Distribution of the masses $({M}/{M_{\bigodot}})$]
    {\centering{\includegraphics[width=60mm]{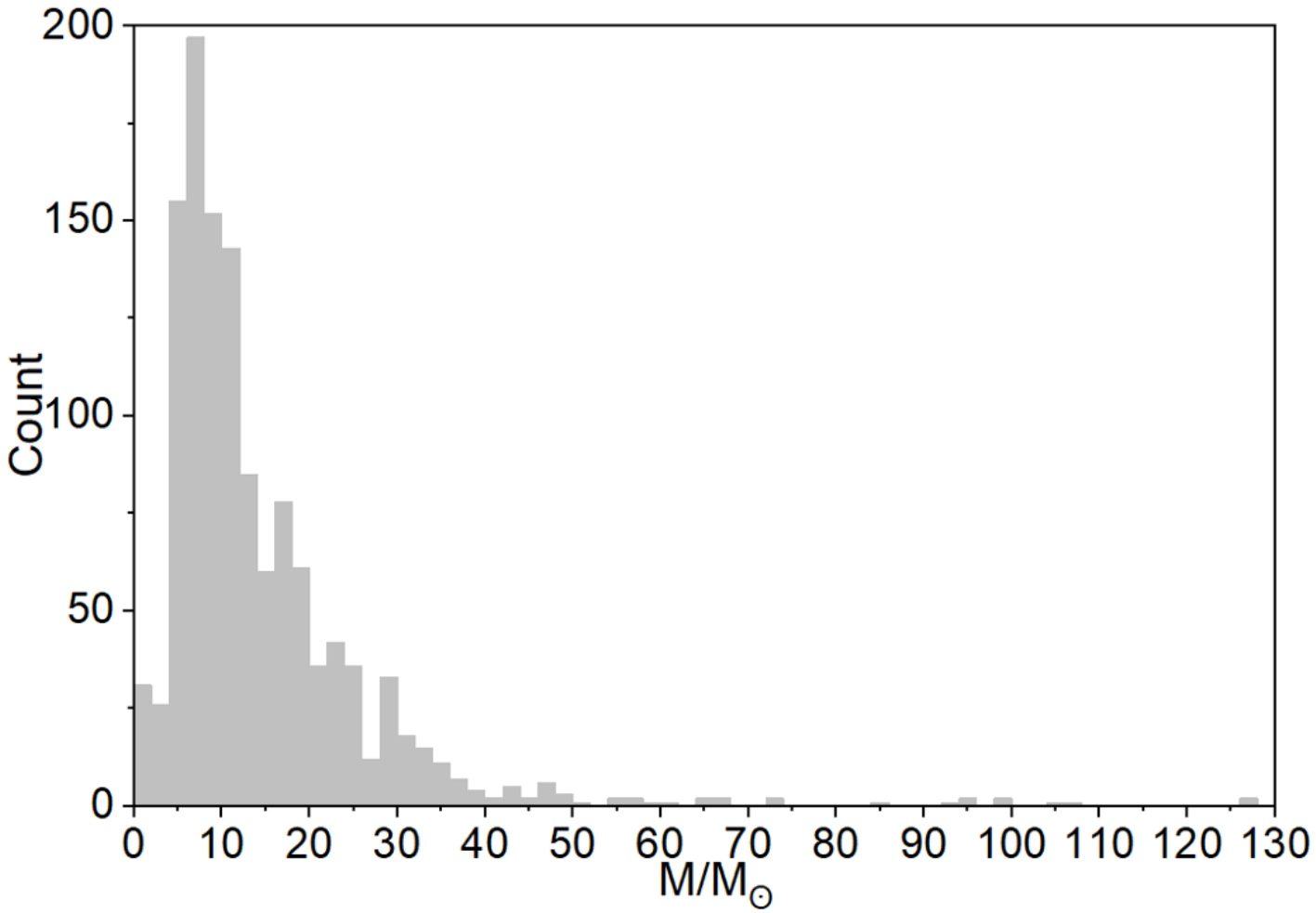}} \label{f10b}}
  \subfigure[Distribution of the temperature $(T_{eff}(K))$]
    {\centering{\includegraphics[width=60mm]{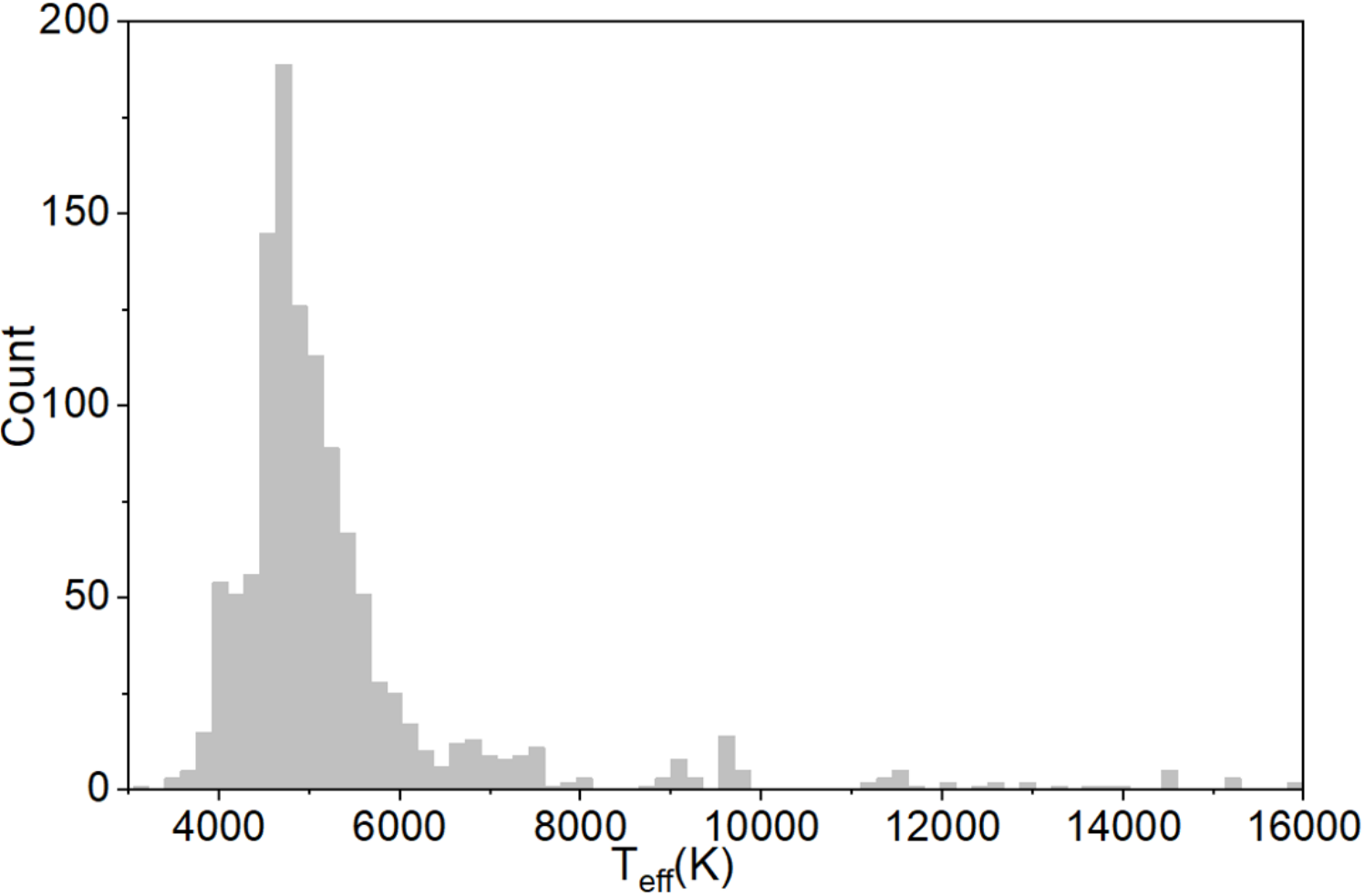}} \label{f10c}}
     \caption[]{Distribution of the stars' radii, masses and temperatures for the open star clusters: NGC 2360.}
    \label{f10}
  \end{center}
\end{figure}

\begin{figure}[!htb]
  \begin{center}
  \subfigure[Distribution of the radii $({R}/{R_{\bigodot}})$]
    {\centering{\includegraphics[width=60mm]{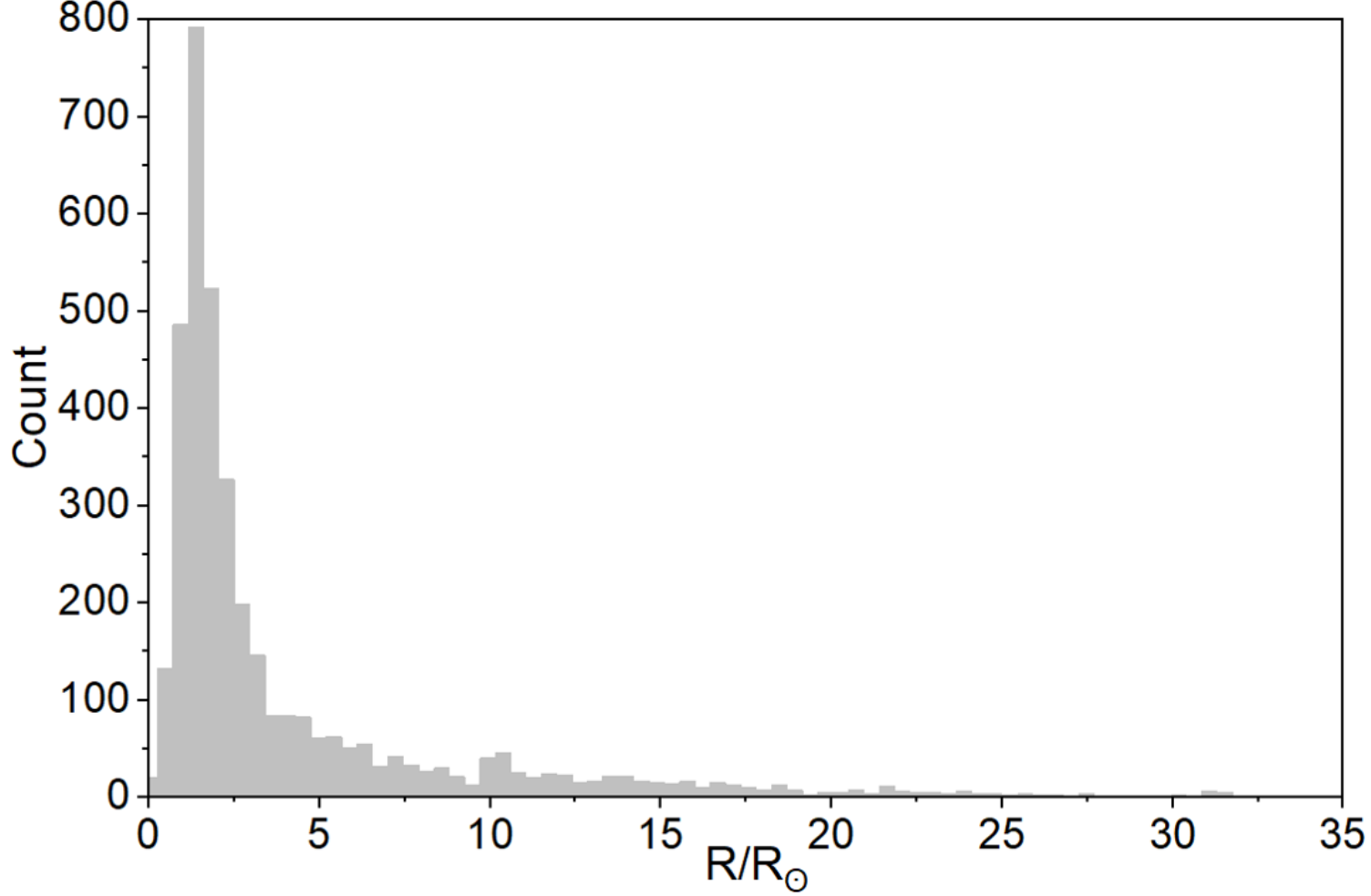}} \label{f11a}}
  \subfigure[Distribution of the masses $({M}/{M_{\bigodot}})$]
    {\centering{\includegraphics[width=60mm]{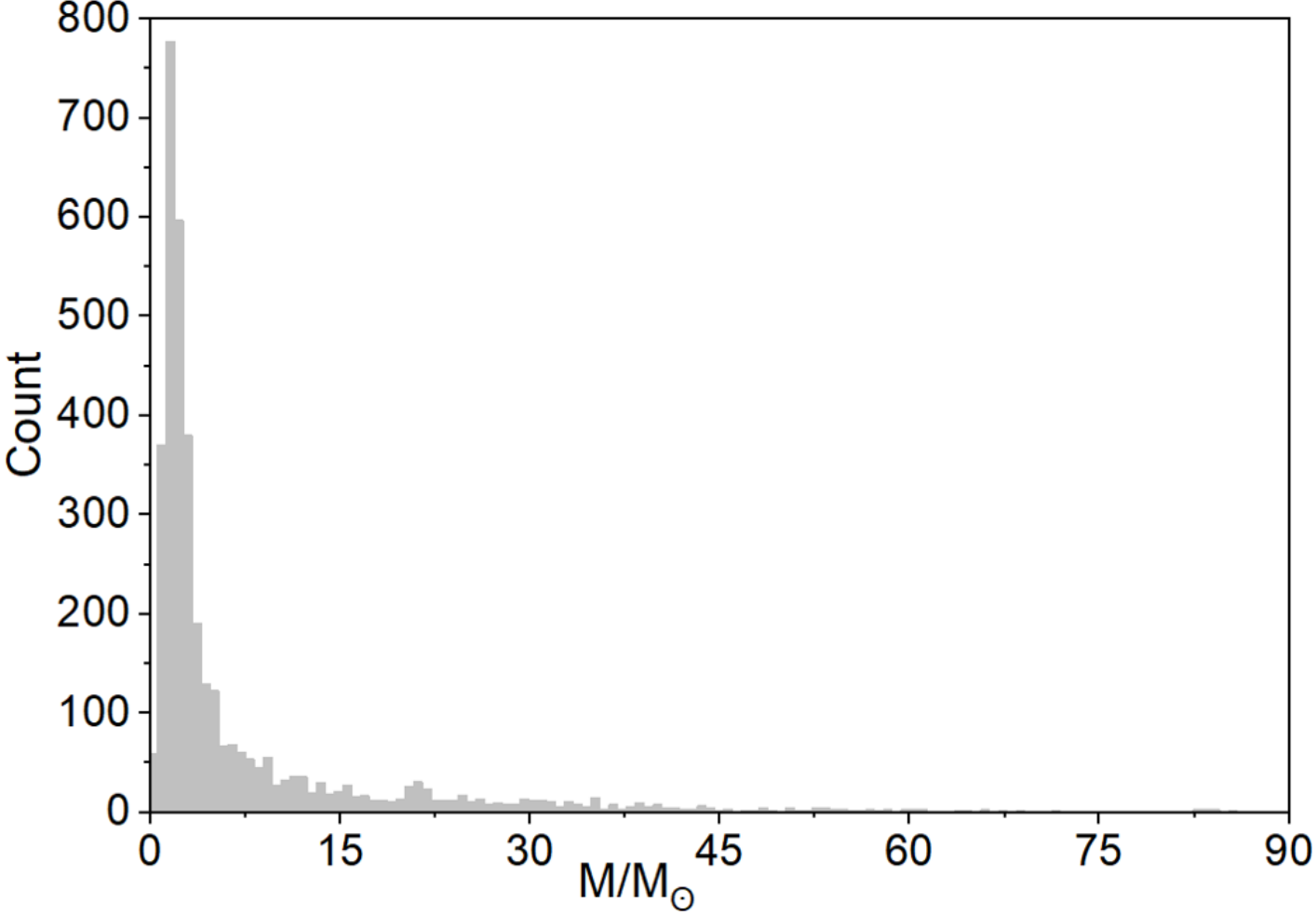}} \label{f11b}}
  \subfigure[Distribution of the temperature $(T_{eff}(K))$]
    {\centering{\includegraphics[width=60mm]{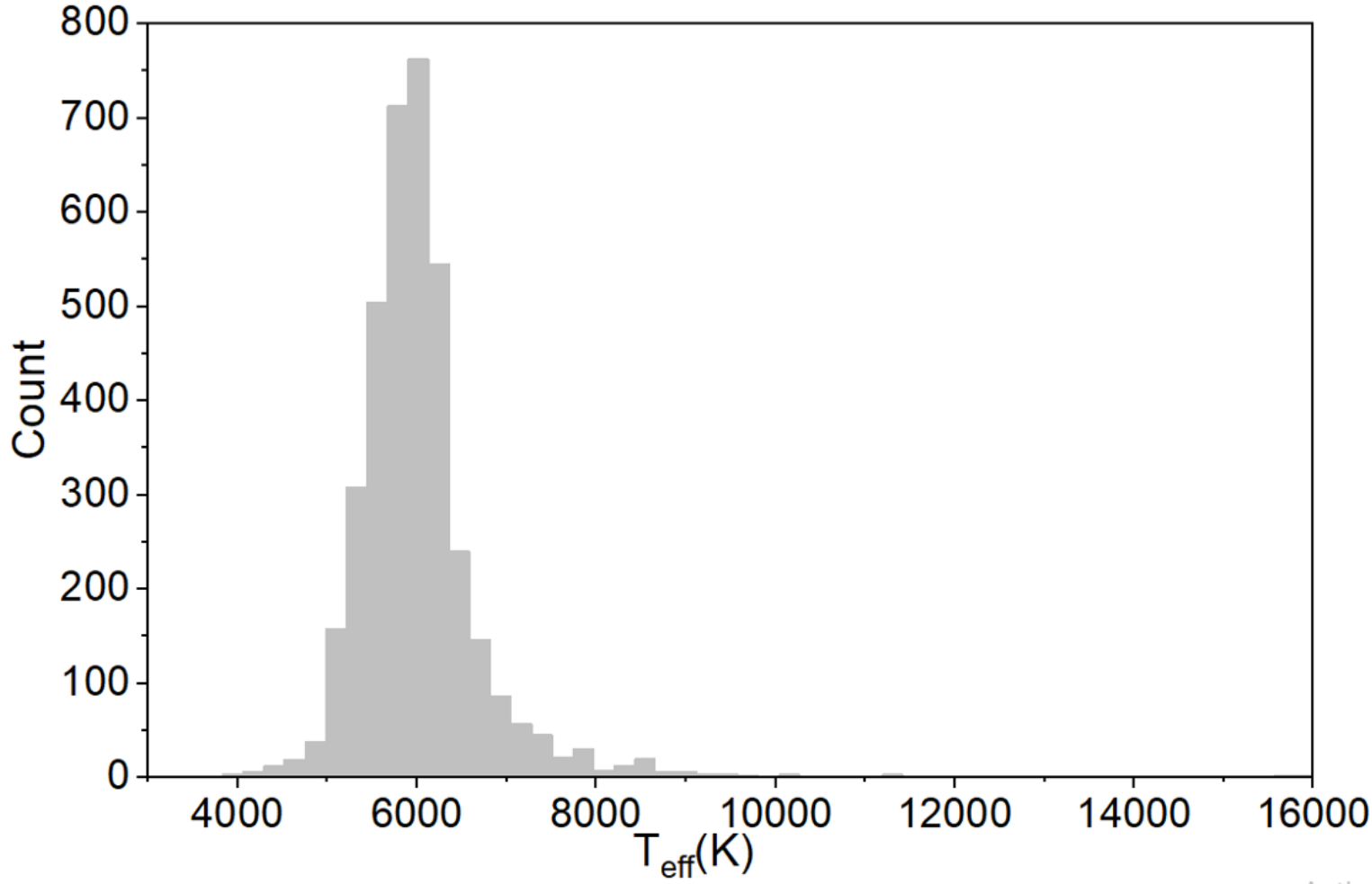}} \label{f11c}}
     \caption[]{Distribution of the stars' radii, masses and temperatures for the globular star clusters: NGC 5272.} \label{f11}
  \end{center}
\end{figure}

In addition to the mass distribution, the aforementioned data for NGC 2360 and NGC 5272 are utilized to obtain the initial mass function (IMF). The IMF depicts the distribution of the stars having an initial mass between $0.1M_{\bigodot}$ and $100M_{\bigodot}$, here $M_{\bigodot}$ corresponds to the solar mass. In other words, the IMF determines the number of stars that have been born with initial masses between $m$ and $m+\Delta(m)$, where $m$ corresponds to the mass. The IMF was first derived by Salpeter \cite{Salpeter} and is popularly known as Salpeter IMF. Further, Kroupa \cite{Kroupa} and Chabrier \cite{Chabrier} have modified the IMF and provided the best known modern mass functions. These IMFs are known as Kroupa IMF and Chabrier IMF, respectively. In this work, the Kroupa IMF is utilized to plot the mass function for the two selected star clusters. The Kroupa IMF is defined in Eq. (\ref{r14})  \cite{Lamers}. In these equations, $N(m)dm$ is the initial mass function, it is also denoted by $\xi(m)\Delta(m)$. The IMF graphs for NGC 2360 and NGC 5272 are illustrated in Fig.\,\ref{f12a} and Fig.\,\ref{f12b}. These graphs are plotted by selecting $C_{1}$, $C_{2}$, and $C_{3}$ such that all the power laws are fitted at the points corresponding to $0.08M_{\bigodot}$ and $0.5M_{\bigodot}$ so that $C_{2}=2C_{1}$ and $C_{3}=25C_{1}$ \cite{Lamers}.

\begin{eqnarray}
 \label{r14}
    \begin{aligned}
    N(m)dm & = C_{1}m^{-2.3} \: for \: m>0.5M_{\bigodot}, \\
    N(m)dm & = C_{2}m^{-1.3} \: for \: 0.08M_{\bigodot}<m<0.5M_{\bigodot}, \\
   N(m)dm & = C_{3}m^{-0.3} \: for \: m<0.08M_{\bigodot},
	\end{aligned}
\end{eqnarray}

\begin{figure}[!htb]
  \begin{center}
  \subfigure[NGC 2360]
    {\centering{\includegraphics[width=60mm]{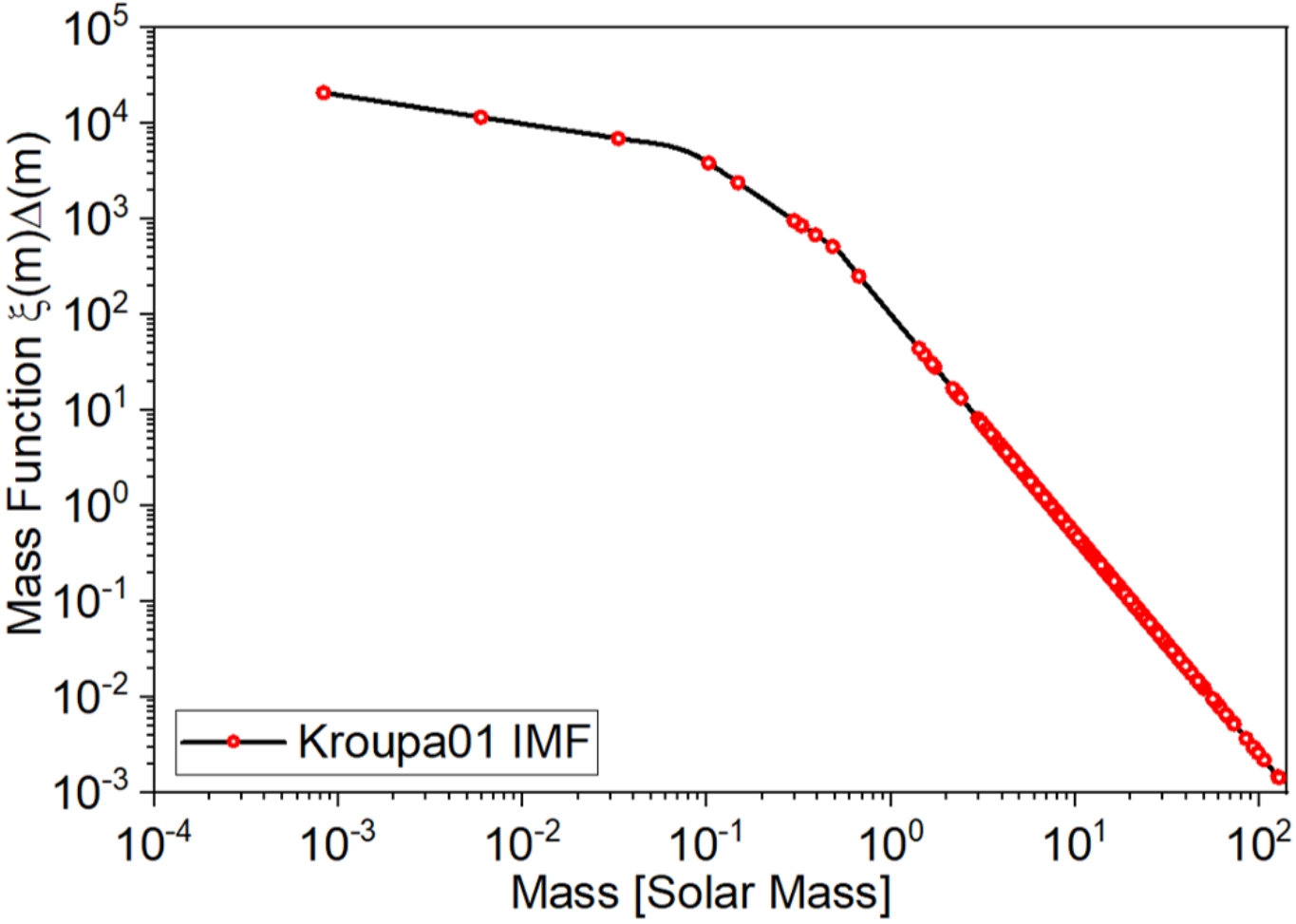}} \label{f12a}}
  \subfigure[NGC 5272]
    {\centering{\includegraphics[width=60mm]{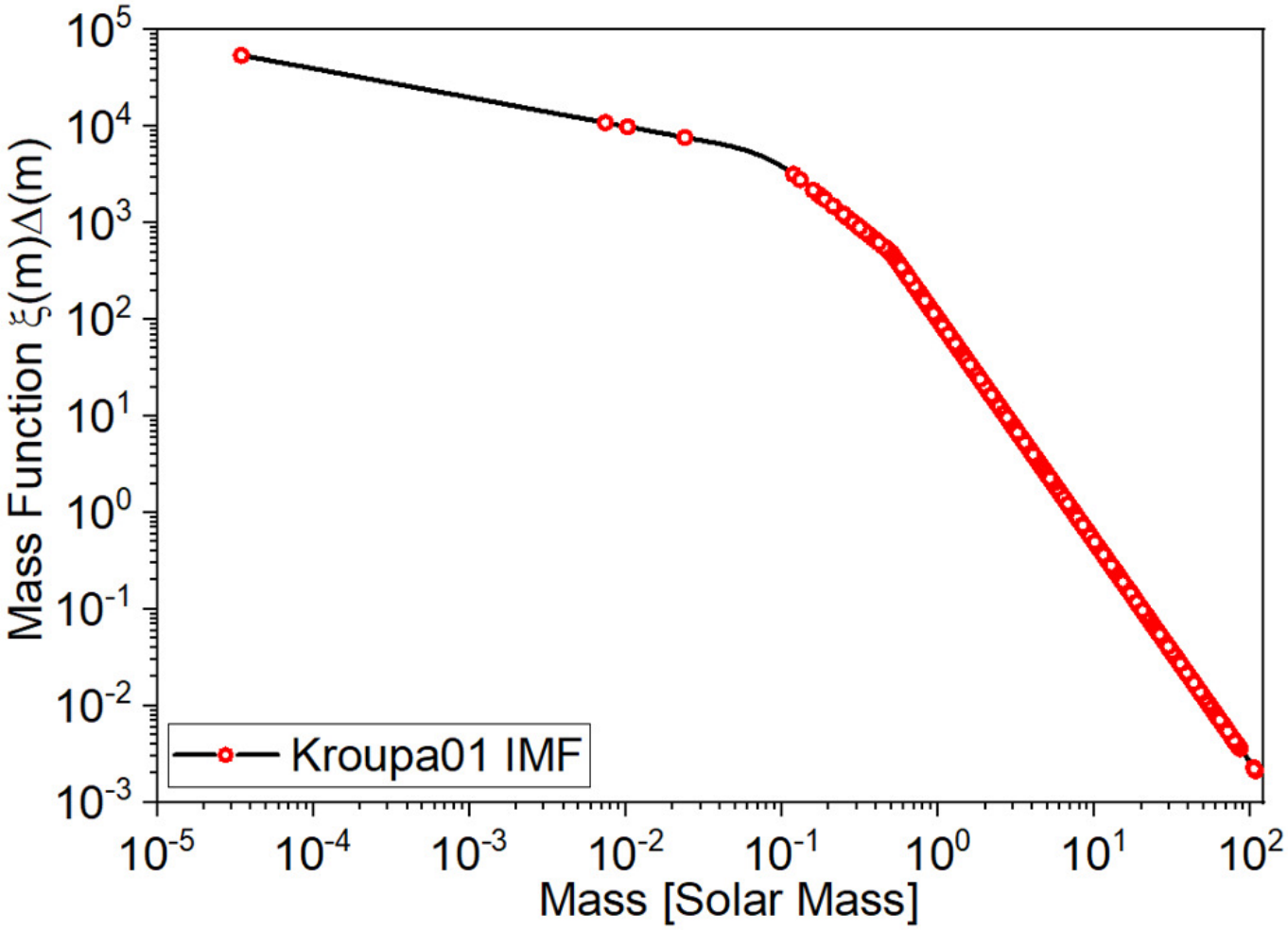}} \label{f12b}}
  \caption[]{Initial mass functions for the selected star cluster using the Kroupa IMF definition.} \label{f12}
  \end{center}
\end{figure}
}

\section*{Conclusion}
\label{sec:Con}

{This work provides insight into the estimation of stellar parameters by manually fitting the best find isochrone on the CMDs of star clusters: NGC 2360 and NGC 5272. The photometric analysis of astronomical images of NGC 2360 and NGC 5272 are performed through a Windows-based software: Aperture Photometry Software. Here, the standard UBVRI photometry system magnitudes are utilized to obtain the color magnitude diagrams (CMDs). These standard  \textit{B}, \textit{V}, and \textit{R} magnitudes are obtained from the \textit{g}, \textit{r}, and \textit{i} calibrated magnitudes. The photometric data utilized are extracted from Sloan Digital Sky Survey (SDSS) data release 12 (DR12) archive. Further, by fitting the isochrones on the CMDs of star clusters, different parameters are extracted. The stellar parameter evaluated are as follows: age of NGC 2360 is found to be 708 Myrs with metallicity, [Fe/H], of -0.15, whereas NGC 5272 is having age of 11.56 Gyrs with metallicity, [Fe/H], of -1.57. In addition, the interstellar reddening, $E(B-V)$, and the distance modulus, $DM$, for NGC 2360 are obtained as 0.12 and 11.65, respectively. While, for NGC 5272, the interstellar reddening is attained as $E(B-V)$=0.015 and distance modulus is $DM$=15.1. The values of these stellar parameters are found to be in close approximation with the previously attained results of literature based on the IRAF analysis technique. Besides these parameters, the other parameters which have been evaluated using photometric data are distance, absorption due to interstellar extinction, temperature, surface brightness, angular radius, luminosity, and initial mass function for the sources present in the two selected star clusters. Thus, this article aids the new entrants in the photometry field to familiarize themselves with the fundamentals of the estimation of the stellar parameter through CMD of star clusters using a windows-based approach. 

Since the core of this work is to provide the basics of photometry analysis selection on the star clusters, their crucial terminologies, and essential parameter estimation by manually  fitting the isochrones (using TOPCAT and EXCEL) on the CMDs. The cleaning of cluster population on the CMDs from the foreground\slash background stars, clearing of spurious objects, error estimations, and the membership determination are not carried out in this work because of a much more sophisticated procedure required for their implementation. These steps would be considered as separate project for analysis.
}

{\textbf{Acknowledgments:}} The authors are grateful to Dr. Rucha Desai (PDPIAS, CHARUSAT), Dr. Dipanjan Dey (ICC, CHARUSAT) and Mr. Parth Bambhaniya (ICC, CHARUSAT) for their valuable suggestions in the completion of this work.


\end{document}